\documentclass[
  aps,
  pra,
  reprint,
  superscriptaddress,
  nofootinbib
]{revtex4-2}
\usepackage{bbm}
\usepackage{soul}
\usepackage{booktabs}
\usepackage{hyperref,xcolor}
\usepackage{physics}
\usepackage{graphicx,amsmath,amsfonts,amssymb}

\newcommand{\uvec}{\boldsymbol{u}}

\newcommand{\omegaTL}{\omega_\mathrm{TL}}

\newcommand{\gammaONEq}{\gamma_{1q}}
\newcommand{\lambdaONEq}{\lambda_{1q}}
\newcommand{\kb}{k_\mathrm{B}}
\definecolor{mycolor}{rgb}{191,0,0}

\begin{document}

\title{Dynamical Regimes of Finite-Length Transmission Lines \\ in Circuit Quantum Electrodynamics}

\author{Fabio Borrelli}
\affiliation{ Department of Electrical Engineering and Information Technology, Universit\`{a} degli Studi di Napoli Federico II, via Claudio 21, Napoli, 80125, Italy}
\author{Giovanni Miano}
\affiliation{Scuola Superiore Meridionale, via Mezzocannone 4,  Napoli, 80125, Italy}
\affiliation{ Department of Electrical Engineering and Information Technology, Universit\`{a} degli Studi di Napoli Federico II, via Claudio 21, Napoli, 80125, Italy}
\author{Adrian Parra-Rodriguez}
\affiliation{Technical University of Munich, TUM School of Natural Sciences, Physics Department, 85748 Garching, Germany}
\affiliation{Walther-Meißner-Institut, Bayerische Akademie der Wissenschaften, 85748 Garching, Germany}
\affiliation{Munich Center for Quantum Science and Technology (MCQST), 80799 Munich, Germany}
\author{Carlo Forestiere}
\email[]{carlo.forestiere@unina.it}
\affiliation{ Department of Electrical Engineering and Information Technology, Universit\`{a} degli Studi di Napoli Federico II, via Claudio 21, Napoli, 80125, Italy}

\begin{abstract}
We study the emergence of continuum, discrete-multimode, and single-mode regimes in finite-length transmission lines capacitively coupled to transmon qubits.  We show that the appropriate description is selected by the hierarchy among the qubit frequency $\omega_q$, the characteristic transmission line frequency $\omega_{\mathrm{TL}}$, and the characteristic coupling frequency $\omega_g$. In the long-line continuum limit, the transmission line acts as a structured reservoir described by a Drude--Lorentz spectral density; in the short-line limit, it reduces to an effective single-mode resonator; and, between these limits, it behaves as a discrete multimode coupler. This provides a unified cQED picture of the dynamical regimes of finite-length transmission lines in superconducting-circuit architectures. 
\end{abstract}

\maketitle

\section{Introduction}
The transmission line paradigm is central to circuit quantum electrodynamics (cQED) \cite{blais_quantum-information_2007,majer_coupling_2007,filipp_multimode_2011,blais_circuit_2021}: superconducting transmission lines can act as coherent quantum busses enabling chip-scale and modular connectivity \cite{kurpiers_deterministic_2018, axline_-demand_2018,leung_speed_2017}; they realize one-dimensional waveguide-QED platforms for superconducting artificial atoms~\cite{abdumalikov_electromagnetically_2010,van_loo_photon-mediated_2013,sheremet_waveguide_2023}; they serve as one-dimensional resonators and provide microwave wiring for control and readout; and when viewed as distributed electromagnetic reservoirs, they form structured environments that can be engineered to tailor open-system dynamics, including strong coupling \cite{ao_extremely_2023}, non-Markovian dynamics \cite{yang_modeling_2026}, collective decay, phase transitions \cite{giacomelli_exact_2026}, entanglement generation and revival \cite{ferreira_collapse_2021}, teleportation \cite{yam_quantum_2026}, and reservoir engineering \cite{paladino_josephson_2003, lalumiere_input-output_2013,kannan_-demand_2023,zanner_coherent_2022}. Transmission lines can therefore play qualitatively different physical roles within the same superconducting-circuit platform, and the appropriate effective description depends on the operating regime.

A substantial body of literature is devoted to the quantization of superconducting networks that combine distributed elements, including transmission lines and more general impedance environments, with lumped circuits. Seminal formulations were given by Yurke and Denker \cite{yurke_quantum_1984} and by Devoret \cite{devoret_quantum_1997}, who established practical quantization rules for microwave circuits. These foundations were subsequently refined and systematized to encompass widely used classes of superconducting quantum circuits \cite{burkard_multilevel_2004,burkard_circuit_2005,vool_introduction_2017,roth_maxwell-schrodinger_2024}, as well as ``black-box'' approaches that quantize complex electromagnetic environments through their normal modes or effective impedance descriptions \cite{nigg_black-box_2012,solgun_blackbox_2014,minev_circuit_2021}. Systems consisting of transmission lines coupled to lumped circuits are characterized by a complete set of commuting observables composed of both discrete and continuous field operators, and various methods have been proposed for their quantization (e.g., \cite{yurke_quantum_1984,devoret_quantum_1997}). Notably, in Ref. \cite{parra-rodriguez_quantum_2018}, by using an auxiliary-mode representation of the  line,  it is shown that for common linear coupling configurations between lumped networks and transmission lines, an intrinsic natural length that provides a high-frequency cutoff emerges. More recently, a Lagrangian formulation was introduced in Ref.~\cite{forestiere_-free_2024} that avoids the use of distributions by enforcing the transmission line boundary conditions  directly within the principle of least action; this formulation was subsequently employed in an input-output approach in Ref.~\cite{forestiere_two-port_2024}. Whereas earlier canonical cQED quantization approaches were restricted to reciprocal networks, Ref.~\cite{parra-rodriguez_exact_2025} generalizes the formalism to transmission lines coupled to one-port and multiport nonreciprocal black-box networks.

Despite these advances, a unified framework is still missing that answers, within a single cQED model, three closely related questions: when can the  line be replaced by a structured continuum environment, when does its  discrete modal structure remain essential, and when is a single-mode  description actually justified? Equally important is the question of which region of circuit parameters supports effectively Markovian dynamics within the continuum description, and which instead gives  rise to substantial memory effects. These questions are particularly relevant for  superconducting-circuit architectures, where finite transmission lines are commonly discussed either in reservoir language or in cavity language, even  though the crossover between these descriptions is experimentally accessible  and can qualitatively reshape the reduced qubit dynamics.

Partial answers to these questions have been obtained in related settings. For example, Refs.~\cite{ao_extremely_2023,ashhab_high-frequency_2024} investigated a qubit coupled to a semi-infinite transmission line, showing that very large multimode Lamb shifts can arise in  strongly coupled  circuit-QED systems. In a complementary context, recent work on Josephson parametric devices coupled to transmission lines has shown that dressed transmission line modes and strongly frequency-dependent coupling can invalidate a Markovian description~\cite{yang_modeling_2026}. These results highlight the importance of going beyond simplified reservoir or single-mode pictures, but they do not provide a unified treatment.

In this work, we introduce a unified dynamical regime map for finite-length transmission lines in cQED.
We consider a transmission line capacitively coupled to two identical transmon qubits. Starting from circuit quantization and employing the auxiliary-mode representation of Ref.~\cite{parra-rodriguez_quantum_2018}, we derive a Hamiltonian in which the transmission line is described as a discrete set of normal modes. Owing to the spatial symmetry of the circuit, these modes separate into even- and odd-parity sectors, which define two independent channels coupled to the qubits through collective operators. This symmetry-resolved decomposition is a key structural simplification of the circuit and provides the basis for the regime analysis developed below.

Our central contribution is to use this cQED Hamiltonian to construct a  unified dynamical regime map controlled by the hierarchy among three  frequency scales: the qubit transition frequency $\omega_q$, the 
transmission line characteristic frequency $\omega_{\mathrm{TL}}$, and the characteristic coupling frequency $\omega_g$, which depends on the characteristic  impedance of the line and on the shunt and qubit-line coupling capacitances. Based on these hierarchies, we identify several operating regions. For each of them, we study the reduced dynamics beyond standard weak-coupling and 
Markovian approximations using the hierarchical equations of motion  (HEOM)~\cite{tanimura_time_1989,tanimura_numerically_2020,
lambert_qutip-bofin_2023}.

We first identify the parameter range in which the discrete transmission line spectrum can be replaced by a continuum, and each parity sector is characterized by an analytic Drude--Lorentz spectral density whose parameters are fixed directly by the circuit elements. This establishes a bridge between the cQED Hamiltonian and the language of open quantum systems, and places the present setup within the broader class of two-qubit systems coupled to a common environment~\cite{reina_extracting_2014,ma_entanglement_2012,saito_dissipative_2007,sinayskiy_non-markovian_2009}. In contrast to phenomenological bath models, however, the environmental spectral density and the associated memory time emerge here explicitly from the circuit parameters. Within this continuum limit, we characterize memory effects through the Breuer--Laine--Piilo (BLP) measure of non-Markovianity~\cite{breuer_measure_2009,clos_quantification_2012}, thereby subdividing the continuum region into subregions with strong and weak information backflow from the environment. Moreover, we show that weak information backflow does not necessarily imply quantitative agreement with perturbative Markovian or weak-memory treatments, such as the secular Gorini--Kossakowski--Lindblad--Sudarshan (GKLS) limit~\cite{gorini_completely_1976,lindblad_generators_1976} and the time-convolutionless (TCL) approach~\cite{shibata_generalized_1977,chaturvedi_time-convolutionless_1979,shibata_expansion_1980,breuer_theory_2002}.
We then study and identify the parameter region in which the continuum approximation breaks down. Here, the dynamics is intrinsically multimode and depends strongly on whether the qubits are resonant or dispersive with respect to the nearby transmission line modes. Finally, we study the short-line limit in which the spectrum is sparse on the scale of the qubit--line hybridization and the dynamics is dominated by one isolated mode, giving rise to cavity-like dynamics.

The paper is organized as follows. In Sec.~\ref{sec:TheModel}, we introduce the Hamiltonian cQED model. In Sec.~\ref{sec:DynamicalRegimes}, we use it to identify the different dynamical behaviors of the system. In Sec.~\ref{sec:Results}, we investigate each of these cases in detail. Finally, in Sec.~\ref{sec:Conclusions}, we present our conclusions.


\section{circuit QED Model}
\label{sec:TheModel}

\begin{figure*}
    \centering
 \includegraphics[width=1\linewidth]{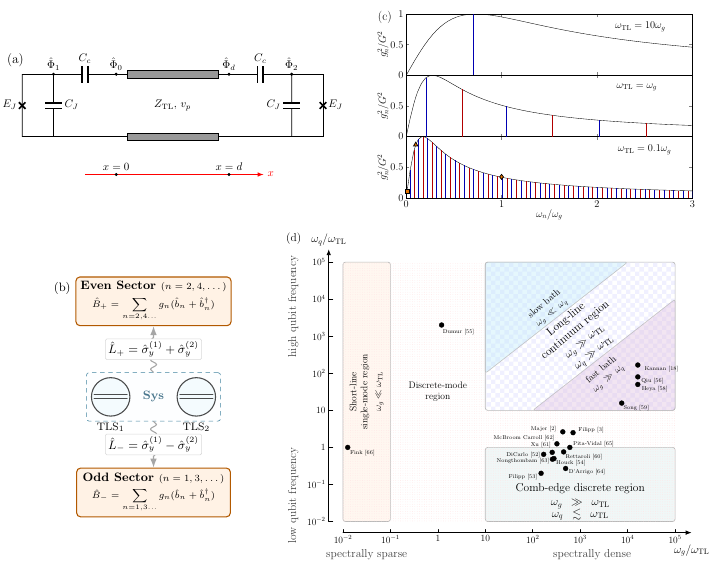}
    \caption{Overview of the finite-length transmission line circuit and of the corresponding
dynamical classification. (a) A transmission line (TL) of length $d$ capacitively coupled through identical
capacitances $C_c$ to two identical transmon qubits with
shunt capacitance $C_J$ and Josephson energy $E_J$. The TL is characterized by impedance $Z_{\mathrm{TL}}$
and propagation velocity $v_p$. (b) Symmetry-resolved representation of the qubit--TL coupling: the TL modes decompose into two independent sectors (even/odd parity), which couple independently to the two qubits through the collective operators $L_{\pm}=\hat{\sigma}_y^{(1)}\pm\hat{\sigma}_y^{(2)}$. (c) Normalized squared coupling strengths $g_n^2/G^2$ of the odd (red line) and even (blue line) sectors versus the normalized mode frequency $\omega_n/\omega_g$ for three values of the TL characteristic frequency: (top) $\omega_{\mathrm{TL}} = 10\,\omega_g$, (center) $\omega_{\mathrm{TL}} = \omega_g$, and (bottom) $\omega_{\mathrm{TL}} = 0.1 \omega_g$. In the bottom plot, the markers denote three representative qubit frequencies: $\omega_q = 0.1\,\omega_{\mathrm{TL}}$ (square), $\omega_q = \omega_{\mathrm{TL}}$ (triangle), and $\omega_q = 10\,\omega_{\mathrm{TL}}$ (diamond). (d) Dynamical map in the plane spanned by $\omega_g/\omega_{\mathrm{TL}}$ and
$\omega_q/\omega_{\mathrm{TL}}$. The markers indicate the position of representative experiments that fit the circuit in panel (a). The order of magnitude of \(\omega_g/\omega_{\mathrm{TL}}\) and \(\omega_q/\omega_{\mathrm{TL}}\) have been extracted from the corresponding references as explained in Appendix \ref{sec:LiteraturePlacement}.}
    \label{fig:finiteTLoverview}
\end{figure*}

We consider the circuit in Fig.~\ref{fig:finiteTLoverview} (a), consisting of a transmission line (TL) of length \(d\) capacitively coupled to two transmon qubits. The qubits (labeled \(i=1,2\)) have identical shunt capacitance \(C_J\) and Josephson energy \(E_J\).  The transmission line is characterized by per-unit-length capacitance \(c\) and inductance \(\ell\); its characteristic impedance is \(Z_{\mathrm{TL}}=\sqrt{\ell/c}\) (and the corresponding propagation velocity is \(v_p=1/\sqrt{\ell c}\)). Each qubit is connected to the line through an identical coupling capacitance \(C_c\), so that the coupling is symmetric under the exchange of the two qubits. The circuit Hamiltonian is (see also Appendices \ref{app:Normalization}, \ref{app:ClassicalHam})
\begin{equation}
\hat{H}=\hat{H}_{t_1}+\hat{H}_{t_2}+\hat{H}_{\mathrm{TL}}+\hat{H}_{\mathrm{int}},
    \label{eq:QuantizedHamiltonian}
\end{equation}
where $\hat{H}_{t_1}$ and $\hat{H}_{t_2}$ are the Hamiltonians of the two transmon qubits, $\hat{H}_{\mathrm{TL}}$ is the transmission line Hamiltonian, and $\hat{H}_{\mathrm{int}}$ is the interaction Hamiltonian between the qubits and the transmission line.

\subsection{Transmon qubits}

The Hamiltonians of the transmon qubits read 
\begin{equation}
\label{eq:CPB1}
\hat{H}_{t_i} =
\frac{\hat{Q}_i^{2}}{2C_J}
- E_{J}\cos\!\left(\frac{2\pi\hat{\Phi}_i}{\Phi_Q}\right),
\end{equation}
where $\hat{Q}_i$ and $\hat{\Phi}_i$ are the node charge and flux operators of qubit $i=1,2$, and $\Phi_Q=h/2e$ is the flux quantum. 

Projecting each transmon qubit onto its two lowest eigenstates, we approximate
\begin{equation}
\hat{H}_{t_i}\approx \frac{\hbar\omega_q}{2}\,\hat{\sigma}_z^{(i)},\qquad i=1,2,
\label{eq:Hsys}
\end{equation}
where $\omega_q$ is the qubit transition frequency and $\hat{\sigma}_{\alpha}^{(i)}$ are Pauli operators acting on qubit $i$.

In the qubit subspace we define the charge and Cooper-pair-number transition matrix elements as
\begin{equation}
Q_T \equiv \bra{g^{(i)}}\hat Q_i\ket{e^{(i)}},\quad
n_T \equiv \bra{g^{(i)}}\hat n_i\ket{e^{(i)}}=\frac{Q_T}{2e},
\label{eq:TransitionElements}
\end{equation}
where $\hat n_i=\hat Q_i/(2e)$ and $\ket{g^{(i)}},\ket{e^{(i)}}$ denote the ground and first excited eigenstates of qubit \(i\).
In general, \(Q_T\) depends on the effective Josephson energy \(E_J\).

 By a convenient choice of  phase, we can identify the charge operator in the two-level subspace as
\begin{equation}
 \hat Q_i = Q_T\,\hat\sigma_y^{(i)} ,\qquad i=1,2,
 \label{eq:Q}
\end{equation}
where $ \hat{\sigma}_y^{(i)} =  i ( \hat{\sigma}_+^{(i)} - \hat{\sigma}_-^{(i)})$, $\hat{\sigma}_+^{(i)} = \ket{e^{(i)}}\bra{g^{(i)}}$, $\hat{\sigma}_-^{(i)} = \ket{g^{(i)}}\bra{e^{(i)}}$.

\subsection{Transmission line}

We quantize the transmission line in the auxiliary-mode representation of Ref.~\cite{parra-rodriguez_quantum_2018}; the black-box quantization in Ref.~\cite{solgun_blackbox_2014} provides an equivalent description.

The transmission line Hamiltonian and the interaction Hamiltonian are then expanded in the complete basis $\{u_n\}$ (see Appendix \ref{app:Normalization}) with canonical coordinates
$(\hat{\varphi}_n,\hat{q}_n)$ satisfying $[\hat{\varphi}_n,\hat{q}_m]=i\hbar\,\delta_{nm}$, which naturally removes TL mode-mode couplings, as proposed in Ref. \cite{parra-rodriguez_quantum_2018}:

\begin{equation}
\label{eq:H_TL}
\hat{H}_{\mathrm{TL}}
=\frac{1}{2C_s}\sum_{n=1}^{\infty}\hat{q}_n^{2}
+\frac{C_s}{2}\sum_{n=1}^{\infty}\omega_n^{2}\hat{\varphi}_n^{2},
\end{equation}
where 
\begin{equation}
\label{eq:Cp}
C_s=\frac{C_J\,C_c}{C_J+C_c}
\end{equation}
is the effective capacitance and 
\begin{equation}
\omega_n=v_p k_n.
\label{eq:FrequencySpacing}
\end{equation}
The set $\{k_n\}$ represents the eigenvalues of the auxiliary eigenvalue problem \eqref{eq:EquationEig}-\eqref{eq:BCs}, which are solutions of the characteristic equations:
\begin{subequations}
\label{eq:parity_eigenvalue_equations}
\begin{alignat}{2}
\tan\!\left(\frac{k_n d}{2}\right)
&= -\frac{\omega_{\mathrm{TL}}}{\omega_g}\,k_n d,
&\qquad&
n=2,4,\ldots,
\label{eq:even_eig}
\\
\cot\!\left(\frac{k_n d}{2}\right)
&= +\frac{\omega_{\mathrm{TL}}}{\omega_g}\,k_n d,
&\qquad&
n=1,3,\ldots.
\label{eq:odd_eig}
\end{alignat}
\label{eq:EigTwoQubits}
\end{subequations}
where
\begin{equation}
\omegaTL=2 \pi \frac{v_p}{d}, 
\label{eq:omegaTL}
\end{equation}
is the transmission line characteristic frequency and
\begin{equation}
\omega_g = \frac{2 \pi}{Z_{\mathrm{TL}}\,C_s}
\label{eq:omegag}
\end{equation}
is the coupling characteristic frequency. We also define the time constants $\tau_{g} = 2 \pi / \omega_{g} = Z_{\mathrm{TL}}\,C_s$ and $\tau_{\mathrm{TL}} = 2 \pi / \omega_{\mathrm{TL}} = d/v_p$. With this convention, \(\omega_g=2\pi\omega_{\mathrm{cut}}\), where \(\omega_{\mathrm{cut}}\) is the cutoff frequency used in Ref.~\cite{parra-rodriguez_quantum_2018}.

The basis functions $\{u_n\}$ associated with the eigenvalues $\{k_n\}$ are
\begin{equation}
u_n(x)=
\begin{cases}
\displaystyle
+A_n\,
\frac{\cos\!\left[k_n\left(x-{d}/{2}\right)\right]}{\cos\!\left[k_n {d}/{2} \right]},
& n=2,4,\ldots,
\\[10pt]
\displaystyle
-\,A_n\,
\frac{\sin\!\left[k_n\left(x-{d}/{2}\right)\right]}{\sin\!\left[k_n {d}/{2} \right]},
& n=1,3,\ldots.
\end{cases}
\label{eq:Modes}
\end{equation}
The normalization coefficients $A_n$ in Eqs. \eqref{eq:Modes} are chosen as (see Appendix \ref{app:Normalization}, \ref{app:ClassicalHam})
\begin{equation}
\label{eq:An}
A_n=u_n(0) = \sqrt{\frac{\omega_{\mathrm{TL}}}
{
\omega_\text{TL}
+
{\omega_g}/{2}
+
2 \pi^2 \, {\omega_n^2}/{\omega_g}}}.
\end{equation}
Using the condition \eqref{eq:EigTwoQubits}  in Eq.~\eqref{eq:Modes} yields the simple parity relation
\begin{equation}
u_n(d)=(-1)^n u_n(0).
\label{eq:EvenOdd}
\end{equation}

Introducing bosonic annihilation and creation operators $\hat b_n$ and $\hat b_n^\dagger$
\begin{equation}
[\hat b_n,\hat b_m^\dagger]=\delta_{nm},\qquad [\hat b_n,\hat b_m]=[\hat b_n^\dagger,\hat b_m^\dagger]=0,
\end{equation}
we perform the canonical transformation
\begin{subequations}\label{eq:secondquant_TL_TWOqubit}
\begin{align}
\hat{\varphi}_n &=
 i \sqrt{\frac{\hbar}{2\omega_n C_s}}\left(\hat{b}_n-\hat{b}_n^\dagger\right),\\
\hat{q}_n &=
 \sqrt{\frac{\hbar\omega_n C_s}{2}}\left(\hat{b}_n+\hat{b}_n^\dagger\right).
\end{align}
\label{eq:CanonicalT}
\end{subequations}
Substituting Eq.~\eqref{eq:secondquant_TL_TWOqubit} into the transmission line Hamiltonian~\eqref{eq:H_TL} yields
\begin{equation}\label{eq:H_TL_bosons}
\hat H_{\mathrm{TL}}
=\sum_{n=1}^\infty\hbar\omega_n\left(\hat b_n^\dagger\hat b_n+\frac{1}{2}\right),
\end{equation}
where the additive zero-point term may be dropped since it only shifts the global energy.

\subsection{Transmission line - qubits interaction}

The interaction Hamiltonian reads:
\begin{equation}
\label{eq:Hint}
\hat{H}_{\mathrm{int}}
=
\frac{\hat{Q}_1}{C_J}\sum_{n=1}^{\infty}u_n(0)\,\hat{q}_n
+
\frac{\hat{Q}_2}{C_J}\sum_{n=1}^{\infty}u_n(d)\,\hat{q}_n.
\end{equation}
Therefore, $\hat{H}_{\mathrm{int}}$ can be recast by separating even and odd modes using Eq.~\eqref{eq:EvenOdd} as
\begin{align}
\begin{aligned}
\hat{H}_{\mathrm{int}}
=&\;
\frac{\hat{Q}_1+\hat{Q}_2}{C_J}\sum_{n=2,4,\ldots}^{\infty}u_n(0)\,\hat{q}_n \\
&\;+
\frac{\hat{Q}_1-\hat{Q}_2}{C_J}\sum_{n=1,3,\ldots}^{\infty}u_n(0)\,\hat{q}_n .
\end{aligned}
\label{eq:Hint_evenodd}
\end{align}
Using the canonical transformation \eqref{eq:CanonicalT} and \eqref{eq:Q}, $\hat{H}_{\mathrm{int}}$ is rewritten as
\begin{align}\label{eq:Hint_bosons2}
\begin{aligned}
\hat{H}_{\mathrm{int}}
=&\,
\hbar \hat{L}_+ 
\sum_{n=2,4,\ldots}^{\infty} g_n \left(\hat{b}_n+\hat{b}_n^{\dagger}\right) \\
&\,+\hbar \hat{L}_-
\sum_{n=1,3,\ldots}^\infty g_n \left(\hat{b}_n+\hat{b}_n^{\dagger}\right),
\end{aligned}
\end{align}
where we introduced the two collective system  operators
\begin{equation}
\hat{L}_\pm = \hat{\sigma}_y^{(1)}\pm\hat{\sigma}_y^{(2)},
\end{equation}
the coupling rates
\begin{equation}\label{eq:gn}
g_n = G \sqrt{\frac{\omega_{\mathrm{TL}}\omega_n}
{ (2\pi \, \omega_n/\omega_g)^2+2 \, \omega_{\mathrm{TL}}/\omega_g+1}},
\end{equation}
the dimensionless factor
\begin{equation}
   G =  n_T\sqrt{8\alpha_\mathrm{fs}}\,\frac{C_s}{C_J}\ \sqrt{\frac{Z_{\mathrm{TL}}}{Z_0}},
   \label{eq:FactorG}
\end{equation}
the fine-structure constant $\alpha_\mathrm{fs}=\frac{1}{4\pi\varepsilon_0}\frac{e^2}{\hbar c_0}\simeq 1/137$ and the vacuum impedance $Z_0=\sqrt{\mu_0/\varepsilon_0}\simeq 377 \Omega$.

In the symmetric configuration we are dealing with, the transmission line modes naturally split into two independent parity sectors, as illustrated in Fig.~\ref{fig:finiteTLoverview} (b). We introduce the bath force operators $\hat{B}_{\pm} = \sum_{n\in \, \substack{2,4,\ldots\\ 1,3,\ldots}} g_n 
\left(\hat b_n+\hat b_n^\dagger\right)$. Importantly, because the corresponding wave numbers are determined by transcendental Eqs. \eqref{eq:EigTwoQubits}, the mode frequencies within each parity sector are not, in general, equally spaced.
The two sectors couple to the qubits through the collective operators \(\hat L_{\pm}\). Their discrete  spectral densities are
\begin{equation}
J_{\pm}(\omega)=\pi \sum_{n\in \, \substack{2,4,\ldots\\ 1,3,\ldots}}
 g_n^2 \delta\left(\omega-\omega_n\right) .
 \label{eq:spec_density_conv}
\end{equation}
From now on, we assume that the two parity sectors are initially in equilibrium thermal states \(\rho_{\rm E}\) at temperature $T$. 
 The bath correlation functions are given by
\begin{multline}
\frac{C_\pm(\tau)}{\hbar^2}= 
\frac{1}{\hbar^2}{\rm Tr}_{\rm E}
\left[
\hat B_\pm(\tau)\hat B_\pm(0)\rho_{\rm E}^{(\pm)}
\right]\\
= 
\sum_{n\in \, \substack{2,4,\ldots\\ 1,3,\ldots}}
g_n^2
\left[
\coth\!\left(\frac{\beta\hbar\omega_n}{2}\right)\cos(\omega_n\tau)
-i\sin(\omega_n\tau)
\right]
\label{eq:CorrFunct_gn}
\end{multline}
where $\beta=1/k_B T$ is the transmission line inverse temperature.

\section{Dynamical Regimes}
\label{sec:DynamicalRegimes}

The dynamical regimes of a finite-length TL are determined by the hierarchy among the qubit transition frequency $\omega_q$, the characteristic frequency $\omega_g$, and the TL characteristic frequency $\omega_{\mathrm{TL}}$. Depending on the relative magnitude of these three scales, the TL is perceived either as an effectively continuous environment or as a discrete system.

This interplay is illustrated in Fig.~\ref{fig:finiteTLoverview} (c), where we plot the normalized coupling weights $g_n^2/G^2$ for the even and odd sectors as a function of the normalized mode frequency $\omega_n/\omega_g$ for three representative values of the line mode spacing: (top) $\omega_{\mathrm{TL}} = 10\,\omega_g$, (center) $\omega_{\mathrm{TL}} = \omega_g$, and (bottom) $\omega_{\mathrm{TL}} = 0.1\,\omega_g$. As $\omega_{\mathrm{TL}}$ decreases relative to $\omega_g$, the modal comb becomes progressively denser. However, a large density of modes alone is not sufficient to justify a continuum description: what also matters is whether the qubit frequency $\omega_q$ probes a region containing many modes on the scale relevant to the interaction.
To make this point explicit in the bottom plot, we also indicate three representative qubit frequencies with different markers: $\omega_q = 0.1\,\omega_{\mathrm{TL}}$, $\omega_q = \omega_{\mathrm{TL}}$, and $\omega_q = 10\,\omega_{\mathrm{TL}}$. The markers show that the same physical TL can be experienced in qualitatively different ways depending on the location of $\omega_q$. When $\omega_q \gg \omega_{\mathrm{TL}}$, the qubit samples many closely spaced modes, and the line is well approximated as a continuum. By contrast, when $\omega_q \lesssim  \omega_{\mathrm{TL}}$, the qubit probes the low frequency edge of the TL spectrum, where the discrete nature of the line remains essential. Thus, even when $\omega_{\mathrm{TL}} \ll \omega_g$, where the modes appear dense on the scale of the coupling envelope, a qubit tuned near the spectral edge still experiences the TL as a sparse few-mode system rather than as a continuum.

In Fig.~\ref{fig:finiteTLoverview}(d), we identify the main dynamical regimes in the parameter plane
\((\omega_g/\omega_{\mathrm{TL}},\omega_q/\omega_{\mathrm{TL}})\). They distinguish when a finite-length transmission line is more naturally described as a continuum reservoir, a discrete multimode coupler, or an effective single-mode. In the following sections, we analyze these three regimes in detail and derive the corresponding effective descriptions. 
This parameter plane provides a natural framework to relate experiments fitting the circuit of panel (a) to the proposed regime map. The positions of the markers should be understood as order-of-magnitude estimates since the quantities reported in the literature are not always sufficient to reconstruct all the parameters with high precision; further details are given in Appendix~\ref{sec:LiteraturePlacement}.
This figure highlights that superconducting-circuit architectures operate in different regions of the parameter plane $(\omega_g/\omega_{\mathrm{TL}},\omega_q/\omega_{\mathrm{TL}})$ or on their borders, motivating a unified cQED treatment.

\subsection{Long-line continuum region}
In the limit $\omegaTL \ll \omega_g$, the right-hand side of Eq.~\eqref{eq:EigTwoQubits} is small, and the even-sector roots approach those of the equation $\tan(k_nd/2)=0$, while the odd-sector roots approach those of $\cot(k_nd/2)=0$. Both can then be approximated by mode angular frequencies
\begin{equation}
\omega_n=v_p k_n=v_p \frac{n \pi}{d}=n \frac{\omega_{\mathrm{TL}}}{2}, \quad n=1,2,3, \ldots.
\label{eq:EquiEigenF}
\end{equation}
Here, the full comb has spacing $\omega_{\mathrm{TL}}/2$, while each individual
parity sector has spacing $\omega_{\mathrm{TL}}$.

The finite-length TL can be treated as a continuum when both of the following conditions hold:
\begin{subequations}
    \label{eq:LongLineConditions}
    \begin{align}
        \label{eq:TLg}
        \omega_{\mathrm{TL}} \ll \omega_g,\\
        \label{eq:TLq}
        \omega_{\mathrm{TL}} \ll \omega_q.
    \end{align}
\end{subequations}
We refer to this parameter region as the  \textit{long-line continuum region} that occupies the upper-right region of the diagram in Fig.~\ref{fig:finiteTLoverview} (d).
These inequalities encode two distinct requirements. 
Condition~\eqref{eq:TLg} states that the spacing between adjacent TL resonances is much smaller than the frequency scale over which the coupling  varies appreciably. 
Condition~\eqref{eq:TLq} ensures that the qubit interacts with a dense set of line modes around its transition frequency $\omega_q$, and it is located well inside the TL mode comb, rather than close to its lower comb-edge. In the continuum limit $\omega_{\mathrm{TL}}\rightarrow 0$ the discrete spectral densities of the two sectors $J_{\pm}(\omega)$ tend to the same continuous function $J(\omega)$, which has the Drude--Lorentz form.

Under these conditions, the TL can be replaced, up to finite-size corrections, by two independent bosonic baths with the same spectral density 
\begin{equation}\label{eq:Jomega_DL}
J(\omega)=\frac{2\lambda\gamma\,\omega}{\gamma^2+\omega^2}.
\end{equation}
The \textit{bath relaxation rate} $\gamma$ and the \textit{bath reorganization rate} $\lambda$ are given by
\begin{subequations}\label{eq:lambda_gamma}
\begin{align}
\label{eq:gamma}
\gamma &=
 \frac{\omega_g}{2\pi}\sqrt{1+2\frac{\omega_{\mathrm{TL}}}{\omega_g}},\\
 \label{eq:lambda}
\lambda &=
\frac{1}{4} G^2
\frac{\omega_g}{\sqrt{1+2\dfrac{\omega_{\mathrm{TL}}}{\omega_g}}}.
\end{align}
\end{subequations}
Under condition \eqref{eq:TLg}, Eq.~\eqref{eq:lambda_gamma} reduces to
\begin{equation}
\gamma   \approx  \frac{\omega_g}{2\pi}; \qquad
\lambda \approx  \frac{\pi}{2} G^2
\gamma . 
\label{eq:LongLimit}
\end{equation}
The correlation functions of the two bosonic baths are equal and are given by
\begin{equation}
\label{eq:CorrFunct_Cont}
\frac{C(\tau)}{\hbar^2} =  \int_0^{\infty}\!d\omega \, \frac{J(\omega)}{\pi}
\left[\coth\!\left(\frac{\beta\hbar\omega}{2}\right)\cos(\omega\tau)-i\sin(\omega\tau)\right].
\end{equation}

\begin{figure*}
    \centering
\includegraphics[width=0.95\linewidth]{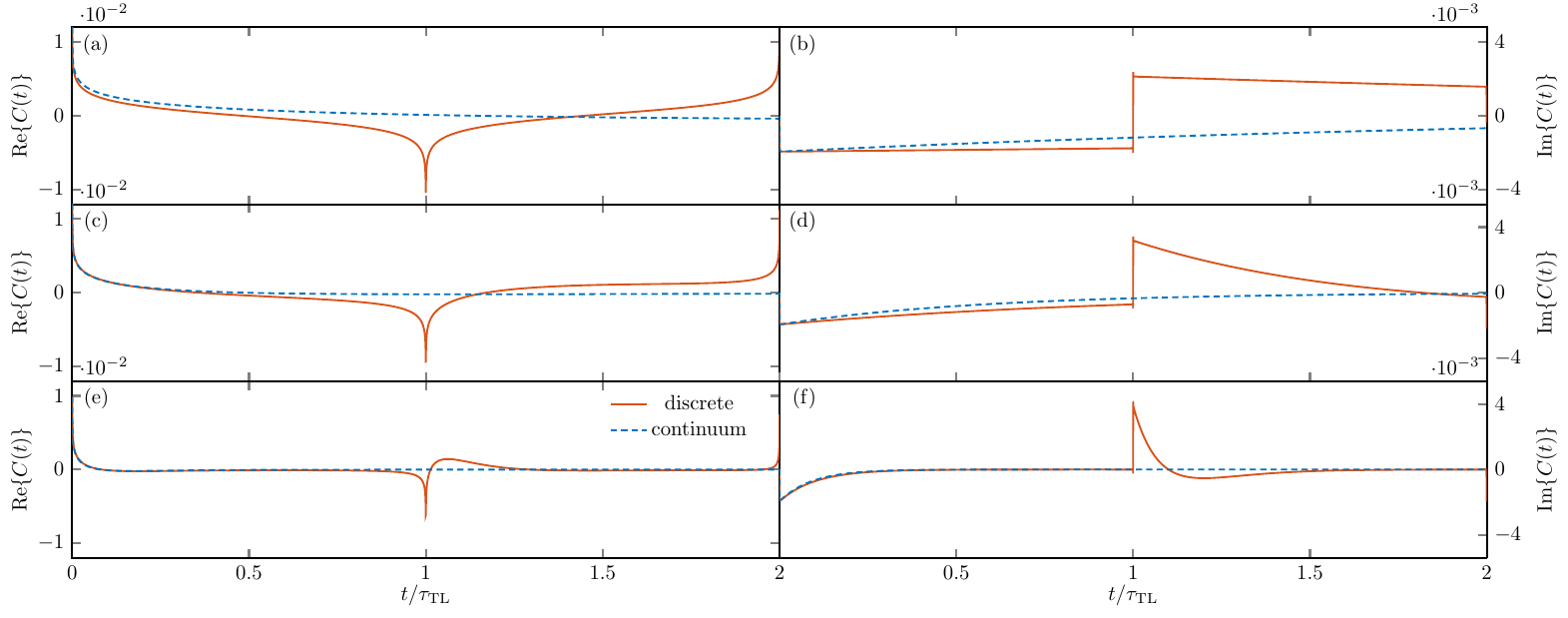}
\caption{Real and imaginary parts of the correlation function $C(t)$ at $T=0$ for the discrete [Eq.~\eqref{eq:CorrFunct_gn}] and continuum [Eq.~\eqref{eq:CorrFunct_Cont}] models. The left column, panels (a), (c), and (e), shows $\mathrm{Re}\{C(t)\}$; the right column, panels (b), (d), and (f), shows $\mathrm{Im}\{C(t)\}$. The three rows correspond to $\omega_{\mathrm{TL}}/\omega_g=10$, $1$, and $0.1$, from top to bottom.}
    \label{fig:CorrelationFunctions}
\end{figure*}

A direct time-domain test of condition \eqref{eq:TLg} is provided by Fig.~\ref{fig:CorrelationFunctions}, where the exact discrete correlation function from Eq.~\eqref{eq:CorrFunct_gn} at zero temperature is compared with the continuum expression~\eqref{eq:CorrFunct_Cont} for three values of $\omega_{\mathrm{TL}}/\omega_g$. The agreement improves systematically as $\omega_{\mathrm{TL}}$ is reduced. In particular, only for $\omega_{\mathrm{TL}} = 0.1\,\omega_g$ do the discrete and
continuum expressions agree over the initial decay window, namely the time interval most relevant for the onset of the reduced qubit dynamics.   
This figure also highlights the intrinsic limitation of the continuum description. The exact discrete correlation function displays revivals at times set by the inverse mode spacing, which reflect the finite propagation time across the line and the recurrence of phase coherence in a finite system.  By construction, such revivals are absent in the continuum approximation, which therefore captures only the pre-revival dynamics.

Since the lowest TL mode occurs at frequency 
$\omegaTL/2$ for the odd sector and $\omegaTL$ for the even sector, 
evaluating the bath correlation function in the continuum 
approximation introduces a temperature-dependent error $\varepsilon_C(\tau)$ coming from 
the spectral region $[0, \omegaTL/2]$ (odd) and $[0, \omegaTL]$ (even) where the continuum 
description spuriously fills in modes that are not present in the 
discrete spectrum, which is analyzed in Appendix \ref{sec:TemperatureRole}.

Finally, Eq.~\eqref{eq:Jomega_DL} illustrates how the bath character depends on the characteristic coupling frequency $\omega_g$. The peak of the spectral density is at $\omega_g/(2\pi)$. Thus, for small $\omega_g$, the spectral weight is concentrated at low frequencies, and the bath correlation function decays slowly; this is the \textit{slow-bath} region in which \textit{memory effects} are enhanced.  By contrast, for large $\omega_g$ (large $\gamma$), the spectrum broadens. The correlation function decays rapidly, corresponding to a \textit{fast bath}. The circuit approaches conditions under which a Markovian description becomes accurate.

\subsection{Discrete-mode region}
\label{sec:DiscreteModeRegime}

The preceding analysis shows that the continuum description is valid for $t<\tau_\mathrm{TL}$ only when both conditions~\eqref{eq:LongLineConditions} hold. If either of them is violated, the reduced dynamics is governed by the discrete transmission line modes. We refer to this complementary region of parameter space as the \textit{discrete-mode region}, shown as the dotted region in Fig.~\ref{fig:finiteTLoverview}~(d).

Two limiting cases are particularly simple. In the comb-edge discrete region, discussed in Sec.~\ref{sec:DiscreteSpectrum}, the qubit is tuned near the low-frequency edge of a dense modal comb, and only a few nearby modes participate significantly in the dynamics. In the \textit{short-line region}, discussed in Sec.~\ref{sec:ShortLineRegime}, the spectrum is sparse on the scale of the coupling and, for typical superconducting-circuit parameters (see Appendix~\ref{sec:OrderMagnitude}), a single transmission line mode is usually sufficient for a quantitatively accurate description.
Away from these two limit regions, however, the number of modes that must be retained can be much larger and depends nontrivially on the hierarchy among $\omega_q$, $\omega_g$, and $\omega_{\mathrm{TL}}$, as well as on the value of $G$.

\subsubsection{Comb-edge discrete region}

\label{sec:DiscreteSpectrum}
It is important to emphasize that condition~\eqref{eq:TLg} alone does not suffice to justify a continuum description. This is already apparent in Fig.~\ref{fig:finiteTLoverview} (c) (bottom): even when $\omega_{\mathrm{TL}} \ll \omega_g$, if $\omega_q \lesssim \omega_{\mathrm{TL}}$ the qubit probes the spectral region near the lower comb-edge, where the density of states is not locally uniform and the full discrete-mode Hamiltonian in Eq.~\eqref{eq:Hint_bosons2} must be retained. Nevertheless, since $\omega_{\mathrm{TL}}/\omega_g \ll 1$, the eigenfrequencies are approximately equally spaced and are well described by Eq.~\eqref{eq:EquiEigenF}.

To maintain coherence with the continuum notation introduced above, the expression of the coupling to the $n$th mode given by Eq.~\eqref{eq:gn} may still be written as
\begin{equation}
g_n^2 = \frac{J (\omega_n)}{\pi} \omega_\text{TL}.
\end{equation}
This expression samples the Drude--Lorentz spectral density \eqref{eq:Jomega_DL} at discrete frequencies. Specifically, we define the \textit{comb-edge discrete region} by the conditions
\begin{subequations}
    \label{eq:DiscreteConditions}
    \begin{align}
        \label{eq:DiscreteTLg}
         \omega_g \gg \omega_{\mathrm{TL}} ,\\
        \label{eq:DiscreteTLq}
       \omega_q  \lesssim \omega_{\mathrm{TL}}.
    \end{align}
\end{subequations}
This region corresponds to the lower-right portion of Fig.~\ref{fig:finiteTLoverview}~(d).

In this region, the bath correlation function is a discrete sum of oscillatory terms rather than a short-lived kernel, so memory effects and revivals arise from the finite set of relevant modes. To distinguish the resulting dynamical behaviors, we define the detuning $\Delta_n=\omega_q-\omega_n$ and the dimensionless parameter \cite{filipp_multimode_2011}
\begin{equation}
\eta_n = \frac{g_n}{|\Delta_n|}.
\label{eq:eta}
\end{equation}
This quantity compares the strength of the qubit--mode coupling with the corresponding detuning and therefore measures whether mode $n$ participates perturbatively or resonantly in the dynamics.
If $\eta_n \ll 1$ for all relevant modes, the dynamics is \textit{dispersive}. By contrast, if one or more modes satisfy $\eta_n \gtrsim 1$, the dynamics enters a \textit{resonant} regime.

In the dispersive scenario, the off-resonant
transmission line modes can be eliminated perturbatively by a Schrieffer--Wolff transformation. To second order in the qubit--mode
couplings \(g_n\) one obtains the qubit Hamiltonian
\begin{equation}
\begin{aligned}
\hat H_{\mathrm{eff}}^{\mathrm{disp}}
=&\,
\frac{\hbar}{2}
\left(
\omega_q+\delta\omega_q
\right)
\left(
\hat\sigma_z^{(1)}+\hat\sigma_z^{(2)}
\right)
\\
&+
\hbar J_{12}
\left(
\hat\sigma_+^{(1)}\hat\sigma_-^{(2)}
+
\hat\sigma_-^{(1)}\hat\sigma_+^{(2)}
\right)
+\cdots .
\end{aligned}
\label{eq:EffectiveDispersiveHamiltonianSymmetric}
\end{equation}
Here the ellipsis denotes higher-order corrections, mode-population
dependent dispersive shifts, and qubit counter-rotating terms that
oscillate at frequencies of order \(2\omega_q\). Since the two qubits are
identical and symmetrically coupled to the two ends of the transmission
line, their Lamb shifts are equal. Including both rotating and
counter-rotating virtual processes gives
\begin{equation}
\delta\omega_q
\simeq
2\omega_q
\sum_n
\frac{g_n^2}{\omega_q^2-\omega_n^2}.
\label{eq:LambShiftDiscreteSymmetricCompact}
\end{equation}
By contrast, the coherent exchange depends on the relative sign of the
mode at the two-qubit positions and it is given by
\begin{equation}
J_{12}
\simeq
\sum_n
(-1)^n
\frac{2\omega_n g_n^2}{\omega_q^2-\omega_n^2}.
\label{eq:ExchangeFullSymmetricCompact}
\end{equation}
This expression agrees with Ref. \cite{filipp_multimode_2011}. 
Equations~\eqref{eq:LambShiftDiscreteSymmetricCompact} and
\eqref{eq:ExchangeFullSymmetricCompact} therefore describe two effects
of the same off-resonant line modes. The Lamb shift is local and
contributes with the same sign for the two qubits, whereas the coherent exchange
is parity weighted. Thus, in the dispersive discrete-mode region, the
qubit–qubit coupling is not a direct instantaneous interaction; rather,  it is
mediated by off-resonant line modes. Equivalent expressions can
also be obtained from the effective impedance or admittance matrix
\cite{solgun_simple_2019,labarca_toolbox_2024}.

\subsubsection{Short-line single-mode region}
\label{sec:ShortLineRegime}
We now consider the limit in which the separation between adjacent modes is much greater than the coupling frequency $\omega_g$, i.e.  
\begin{equation}
\omega_g \ll  \omega_{\mathrm{TL}}.
\label{eq:ShortLine}
\end{equation}
This region corresponds to the left-hand side of Fig.~\ref{fig:finiteTLoverview} (d). As we have seen from Fig.~\ref{fig:finiteTLoverview} (c), in this case a continuum description of the spectral density fails to describe the correlation function irrespective of the qubit frequency. The dynamics must instead be described in terms of a discrete set of line modes, and the central question becomes how many of them must be retained to obtain a quantitatively reliable description. In this limit, the  expression of $g_n$, given by Eq.~\eqref{eq:gn}, reduces to 
\begin{equation}
g_n\simeq
\begin{cases}
\dfrac{G \omega_g}{2\sqrt{\pi}}
\left(
\dfrac{\omega_{\mathrm{TL}}}{2\omega_g}
\right)^{1/4},
& n=1, \\[4mm]
\dfrac{G\omega_g}{\pi\sqrt{2(n-1)}},
& n=2,3,\ldots .
\end{cases}
\end{equation}
For realistic circuit parameters, the prefactor $G$ is less than one (see Appendix~\ref{sec:OrderMagnitude}), so that the coupling rate is typically much smaller than the TL mode spacing $g_n \ll \omega_{\mathrm{TL}}$.

This hierarchy has two important consequences. First, neighboring TL modes remain well resolved on the scale of the coupling frequency, which makes multimode interaction difficult. Second, the coupling decreases with mode index as $g_n \propto 1/\sqrt{n}$; together with the increasing detuning from higher modes, this provides an effective ultraviolet suppression of far-off-resonant modes \cite{parra-rodriguez_quantum_2018}. Therefore, in the short-line region, the dynamics is well approximated by retaining only the mode whose frequency $\omega_n$ is closest to the qubit frequency $\omega_q$.

\section{Qubit time evolution}
\label{sec:Results}

We now analyze the time evolution of the reduced density operator of the two qubits in the regions of the parameter plane
\((\omega_g/\omega_{\mathrm{TL}},\omega_q/\omega_{\mathrm{TL}})\) introduced in the previous section. 

In the long-line continuum region (Sec.~\ref{sec:LongLine}), where the TL acts as an effective structured bosonic bath, we study the non-Markovianity of the reduced dynamics. 

In the comb-edge discrete region (Sec.~\ref{sec:DiscreteRegime}), where only a finite number of TL modes contribute significantly to the dynamics, we distinguish dispersive from resonant multimode behavior. 

Finally, in the short-line region (Sec.~\ref{sec:ShortRegime}) the dynamics is dominated by an individual TL mode.  

All simulations were performed with the hierarchical equations of motion (HEOM) method~\cite{lambert_qutip-bofin_2023} summarized in Appendix~\ref{sec:HEOMCorrelationExpansion}. We used the QuTiP–BoFiN implementation~\cite{lambert_qutip-bofin_2023} built on QuTiP \cite{johansson_qutip_2012}. Specifically, the hierarchy solved in the simulations is given by the system of first order differential equations presented in Eq.~\eqref{eq:HEOMeq}, and the numerical parameters and convergence checks are reported in Appendix~\ref{sec:HEOMCorrelationExpansion}.

\subsection{Long-line continuum region}
\label{sec:LongLine}

The TL can be treated as a continuum environment when both conditions in Eq.~\eqref{eq:LongLineConditions} are satisfied. To develop physical intuition for the case of two qubits symmetrically coupled through a TL, we begin with the simpler setting of a single qubit coupled to a short-circuited TL described in Appendix \ref{sec:LineShort}. This allows us to isolate the basic mechanisms associated with the structured bosonic bath before addressing the two-qubit problem.  It also provides a useful reference point for assessing the validity of Markovian approximations. In this respect, related recent work has shown, in the complementary context of Josephson parametric devices coupled to transmission lines, that dressed
transmission line modes and frequency-dependent coupling can invalidate a Markovian description~\cite{yang_modeling_2026}.

\subsubsection{A short-circuited TL coupled to a single qubit}

We now consider a short-circuited TL coupled to a qubit. In the long-line continuum
region, the line can be described by a single bosonic bath with a Drude–Lorentz spectral density (given by Eq.~\eqref{eq:Jomega_DL}), a bath relaxation rate $\gammaONEq$ (given by Eq.~\eqref{eq:GammaLong_1Q}), while the reorganization rate $\lambdaONEq$ (given by Eq.~\eqref{eq:LambdaLong_1Q}) is kept fixed, i.e., $\lambdaONEq = 0.1\, \omega_q$. 

Figure~\ref{fig:LongLineContinuum} (a) shows the Breuer--Laine--Piilo non-Markovianity measure $\mathcal{N}(\Phi)$ \cite{clos_quantification_2012,breuer_measure_2009} for the reduced density operator of the qubit. The BLP measures the information backflow from the bosonic bath to the qubit. In Appendix \ref{sec:BLP} we have provided a short introduction to BLP. 

As expected from general analyses of the spin-boson model \cite{clos_quantification_2012}, the dynamics is strongly non-Markovian for a small bath relaxation rate $\gammaONEq/\omega_q$, corresponding to large time constants $\tau_g = Z_\mathrm{TL} C_s$ compared with $1/\omega_q$; the BLP increases significantly as the temperature rises. The bosonic bath is \textit{slow} and retains correlations on timescales comparable with the intrinsic system dynamics. This shows that thermal fluctuations alone are not sufficient to guarantee Markovian behavior if the bath memory time is still long. 

A remarkable feature of Fig.~\ref{fig:LongLineContinuum} (a) is the dark wedge that appears at an intermediate bath relaxation rate and low temperatures. In agreement with previous studies \cite{clos_quantification_2012}, this structure shows that the dependence of $\mathcal{N}(\Phi)$ on $\gammaONEq/\omega_q$ is non-monotonic: within the non-Markovian region, there exists a narrow parameter window in which information backflow is strongly suppressed. As extensively discussed in Ref. \cite{clos_quantification_2012}, this quasi-Markovian pocket can be understood in terms of a condition between the qubit transition frequency $\omega_q$ and the maximum of the effective temperature-dependent environmental spectral density. The corresponding resonance locus is given by \cite{clos_quantification_2012}
\begin{equation}
\left.\frac{\gammaONEq(T)}{\omega_q}\right|_{\mathrm{res}}= \sqrt{\dfrac{ \dfrac{\kb T}{\hbar\omega_q} \sinh \left(\dfrac{\hbar \omega_q}{\kb T}\right)+1}{\dfrac{\kb T}{\hbar\omega_q} \sinh \left(\dfrac{\hbar\omega_q}{\kb T}\right)-1}},
\label{eq:ResonanceLocus}
\end{equation}
 and shown in Fig.~\ref{fig:LongLineContinuum} (a) with a dashed red curve.
 Near this curve, the local spectral profile sampled by the qubit leads to a strong suppression of the BLP measure (see Appendix \ref{sec:BLP}). In this sense, a Markovian region emerges inside a broader non-Markovian domain.

Finally, for large values of both $\gammaONEq/\omega_q$ and $k_\mathrm{B}T/\hbar\omega_q$, the non-Markovianity measure becomes negligibly small. This is consistent with the broadband, short-memory regime of the bosonic bath, for which a time-local Markovian description provides an accurate approximation of the reduced qubit dynamics.

\subsubsection{A TL coupled to two qubits}

Having established this behavior in the single-qubit case, we now turn to the two-qubit setup. This comparison is useful because it allows us to unravel which features of the BLP landscape are inherited from the structured TL bath and which instead arise from the collective nature of the qubit–bath coupling.

\begin{figure*}
  \centering
    \includegraphics[width=\linewidth]{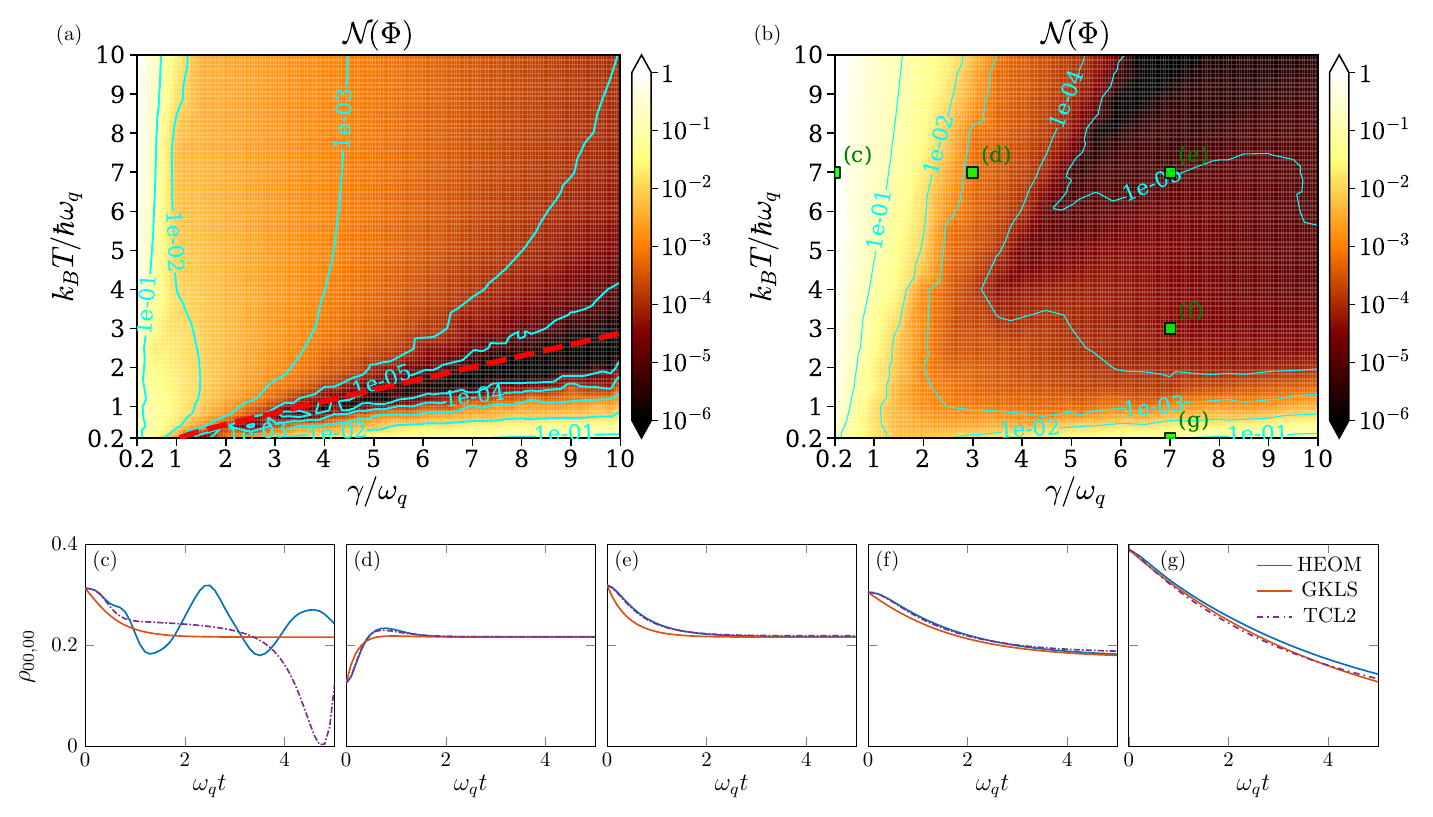}   
\caption{
Long-line continuum region. BLP non-Markovianity measure $\mathcal{N}(\Phi)$ and representative population dynamics in the long-line continuum region. 
Panels (a) and (b) show $\mathcal{N}(\Phi)$ as a function of the normalized bath relaxation rate $\gamma/\omega_q$ and normalized temperature $k_{\mathrm{B}}T/\hbar\omega_q$, for a single-qubit benchmark and for two qubits capacitively coupled to a finite TL, respectively. 
In both cases, the reorganization rate is fixed to $\lambda=0.1\,\omega_q$, and the cyan curves denote contours of constant $\mathcal{N}(\Phi)$. 
In panel (a), the red dashed curve indicates the resonance locus defined by Eq.~\eqref{eq:ResonanceLocus} where $\mathcal{N}(\Phi)$ tends to zero. 
The green squares in panel (b) mark the parameter sets used in the dynamical simulations shown in panels (c)--(g).  The two-qubit initial states used in the BLP maximization are reported in Table~\ref{tab:bloch} of Appendix \ref{sec:BLP}.
Panels (c)--(g) compare HEOM, secular GKLS, and TCL2 predictions for the two-qubit ground-state population $\rho_{00,00}(t)$. 
Panels (c)--(e) correspond to $k_{\mathrm{B}}T/\hbar\omega_q=7$ and $\gamma/\omega_q=0.2$, $3$, and $7$, with $\mathcal{N}(\Phi)=1.61$, $5.6\times10^{-4}$, and $1.11\times10^{-5}$, respectively. 
Panels (f) and (g) correspond to $\gamma/\omega_q=7$ and $k_{\mathrm{B}}T/\hbar\omega_q=3$ and $0.2$, with $\mathcal{N}(\Phi)=4.25\times10^{-5}$ and $7.8\times10^{-2}$, respectively.
}
  \label{fig:LongLineContinuum}
\end{figure*}

Figure~\ref{fig:LongLineContinuum} (b) shows that, in the two-qubit configuration, the BLP non-Markovianity is still primarily controlled by the same two environmental parameters, namely the bath relaxation rate and the temperature, but its detailed structure is now reshaped by the collective coupling to the TL. In particular, the environment couples to the parity operators $\hat{L}_\pm$ so that information backflow  is enhanced, resulting in a stronger effective coupling.  
As in the single-qubit case, the low bath relaxation rate region corresponds to a slow reservoir with long-lived correlations, resulting in a high $\mathcal{N}(\Phi)$, regardless of the temperature. This confirms that, as in the single qubit, thermal fluctuations alone do not restore Markovianity when the environmental memory time is long. However, the two-qubit reduced density operator no longer exhibits the quasi-Markovian pocket characterizing Fig.~\ref{fig:LongLineContinuum} (a). Indeed, the competition between the even and odd-parity channels smooths the crossover between strongly non-Markovian and weakly non-Markovian regions.

Conversely, for intermediate-to-large bath relaxation rates and moderate-to-high temperatures, $\mathcal{N}(\Phi)$ becomes very small over a broad portion of the parameter space. In this parameter region, both collective channels effectively experience a broadband reservoir, and the reduced dynamics approach a time-local, weak-memory description. Overall, compared with the single-qubit case, the two-qubit dynamics is less dominated by a single resonance condition and is instead governed by a broader collective crossover between strong-memory behavior and quasi-Markovian behavior.

Fig.~\ref{fig:LongLineContinuum} (c)--(g) reports the time evolution of the ground-state population \(\rho_{00,00}(t)\) at representative points of the BLP map shown in Fig.~\ref{fig:LongLineContinuum} (b) indicated by green square markers. For each parameter set, the initial condition for the reduced density operator is chosen in such a way as to maximize the BLP measure. We compare the time evolution obtained by HEOM with the time evolutions obtained by the time-convolutionless master equation (TCL2) \cite{shibata_generalized_1977,chaturvedi_time-convolutionless_1979,shibata_expansion_1980,breuer_theory_2002} (Eq.~\eqref{eq:TCL2} derived in Appendix~\ref{app:TCL2}) and by the secular  Gorini--Kossakowski--Lindblad--Sudarshan (GKLS) master equation \cite{gorini_completely_1976,lindblad_generators_1976} (Eq.~\eqref{eq:GKLS_master_local} derived in Appendix \ref{app:GKLSeq}).
The three approaches differ in the level of approximation used to treat the bosonic baths. HEOM provides the reference dynamics: once the exponential expansion of the bath correlation functions and the hierarchy depth are converged, it gives a non-perturbative description of the qubit reduced dynamics. TCL2, on the other hand, is perturbative to second order in the system--bath coupling but retains the finite bath memory through time-dependent coefficients. Finally, the GKLS equation is obtained after the additional Born--Markov and secular approximations, leading to a time-local Markovian generator with constant rates. Comparing these three descriptions, therefore, allows us to distinguish genuine memory effects from corrections due to finite system--bath coupling and from the breakdown of the Markovian approximation.

The comparison between HEOM, GKLS, and TCL2 is performed for the parameter sets
$
\left(\gamma/\omega_q,\,\kb T/\hbar\omega_q\right)
=$$ (0.2,7)$, $(3,7)$, $(7,7)$, $(7,3)$, and $(7,0.2)$. We first consider the horizontal cut at fixed temperature \(\kb T/\hbar\omega_q=7\), corresponding to Fig.~\ref{fig:LongLineContinuum} (c-e). Panel~(c), with \(\gamma/\omega_q=0.2\), lies in the strongly non-Markovian region of the BLP diagram. In this case, the HEOM dynamics display pronounced damped oscillations and partial revivals of \(\rho_{00,00} \), which are clear signatures of environmental memory. The GKLS equation fails qualitatively, as it predicts an almost monotonic relaxation and completely misses the revivals. TCL2 performs better at short times, reproducing the initial trend of the HEOM curve, but it becomes unreliable at later times, where it deviates strongly and eventually yields incorrect long-time behavior.

As \(\gamma/\omega_q\) is increased to \(3\) and \(7\) [panels~(d) and~(e)], the BLP measure is strongly suppressed, and the dynamics become much closer to a monotonic relaxation. In panel~(d), corresponding to \((\gamma/\omega_q,\kb T/\hbar\omega_q)=(3,7)\), the three approaches are already very close, with only minor discrepancies during the initial transient. In panel~(e), at \((7,7)\), the agreement further improves: TCL2 is nearly indistinguishable from HEOM over the full time window, while GKLS still shows a small but visible deviation, mainly in the transient  and in the approach to the asymptotic value. Overall, along this high-temperature cut, increasing \(\gamma\) weakens memory effects and progressively restores the validity of perturbative and Markovian descriptions.

We next examine the vertical cut at a fixed bath relaxation rate \(\gamma/\omega_q=7\), shown in Fig.~\ref{fig:LongLineContinuum} (e-g). At \(\kb T/\hbar\omega_q=3\) [panel~(f)], both TCL2 and GKLS reproduce the HEOM dynamics rather well, consistent with the very small BLP value at this point. TCL2 remains slightly more accurate, especially in the transient region.
At the lower temperature $k_{\mathrm{B}}T/\hbar\omega_q=0.2$ [panel~(g)], the situation is more subtle. 
Here the BLP measure, $\mathcal{N}(\Phi)=7.8\times10^{-2}$, is two to nearly four orders of magnitude larger than in panels~(d)--(f), while still being about a factor of $20$ smaller than in the strongly non-Markovian case of Fig.~\ref{fig:LongLineContinuum}(c). 
Despite the absence of pronounced population revivals, the HEOM dynamics show a systematic quantitative deviation from both approximate treatments over the entire evolution. 
In particular, GKLS and TCL2 nearly overlap, but both predict a faster relaxation than HEOM. 

These comparisons highlight two main points. First, in the region of large BLP, the breakdown of GKLS is evident, and even TCL2 can remain reliable only for short times. Second, when the BLP measure is small, TCL2 generally provides the best approximation to HEOM; however, a small non-Markovianity does not necessarily imply quantitative agreement with a GKLS treatment. In particular, the low-temperature case \((\gamma/\omega_q,\kb T/\hbar\omega_q)=(7,0.2)\) shows that even in the presence of weak information backflow, there can be appreciable corrections to GKLS and TCL2 (even in the absence of pronounced oscillations in the population dynamics).

\subsection{Comb-edge discrete region}

\label{sec:DiscreteRegime}

\begin{figure}
  \centering
    \includegraphics[width=\linewidth]{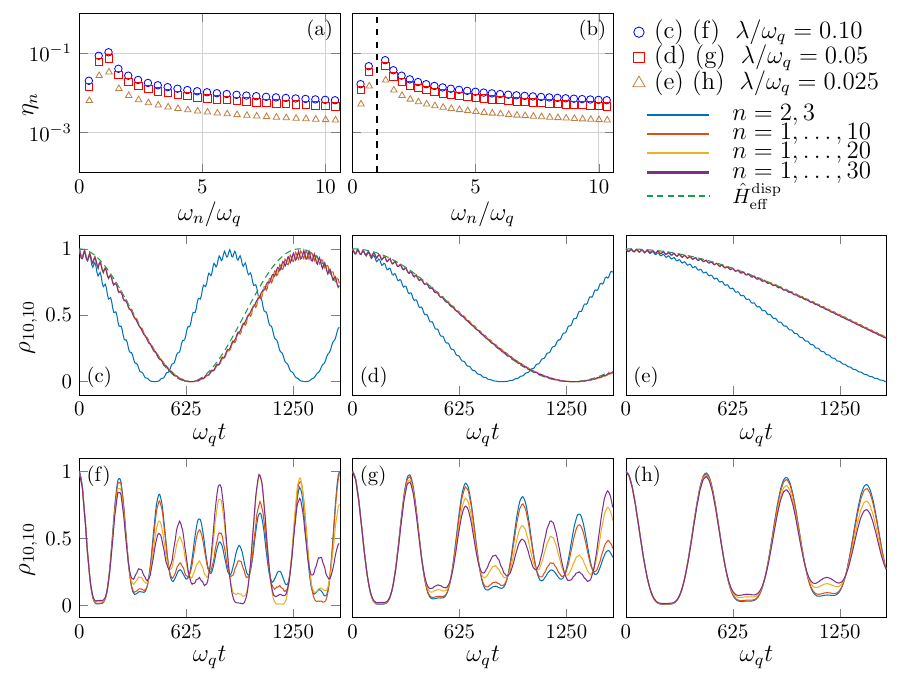}
 \caption{Comb-edge discrete region: mode-resolved coupling parameter $\eta_n=g_n/|\omega_q-\omega_n|$ and two-qubit population dynamics. 
Panels (a) and (b) show $\eta_n$ as a function of the normalized TL mode frequency $\omega_n/\omega_q$ for three reorganization rates, $\lambda/\omega_q=0.10$, $0.05$, and $0.025$, with $\omega_{\mathrm{TL}}/\omega_g=1\times10^{-3}$. The associated dimensionless prefactors are $G=2.24\times10^{-2}$, $1.58\times10^{-2}$, and $1.12\times10^{-2}$. 
Panel (a) corresponds to the dispersive configuration, $\omega_q=(\omega_2+\omega_3)/2$, where the qubit frequency lies between the second and third TL modes. 
Panel (b) corresponds to the resonant configuration, $\omega_q=\omega_3$, where the qubits are resonant with the third TL mode indicated by the vertical black dashed line. 
Panels (c)--(e) show the population $\rho_{10,10}(t)$ for the dispersive configuration shown in (a). The dashed lines show the dynamics generated by the effective dispersive
Hamiltonian of Eq.~\eqref{eq:EffectiveDispersiveHamiltonianSymmetric}, using  $30$ modes. Panels (f)--(h) show the corresponding resonant case in (b). 
In all population plots, the TL is initially in a thermal state at $k_B T/\hbar\omega_q=0.08$, and the initial two-qubit state is $\ket{10}$, corresponding to a single excitation localized on the first qubit. 
The line colors indicate the TL mode indices $n$ retained in the simulations.
}
  \label{fig:CombEdge}
\end{figure}

In this section, we focus on the \emph{comb-edge discrete region} introduced in Sec.~\ref{sec:DiscreteSpectrum}. To this end, we set $\omega_{\mathrm{TL}} = 10^{-3}\,\omega_g$ and choose the qubit frequency $\omega_q$ to be of the same order as $\omega_{\mathrm{TL}}$. Specifically, we compare two representative scenarios: a \emph{dispersive} one, in which the qubit frequency is placed midway between the second and third TL modes, i.e. $\omega_q = (\omega_2 + \omega_3)/2$, and a \textit{resonant} one, in which the qubits are tuned to the third TL mode  $\omega_q = \omega_3$. 

Figure~\ref{fig:CombEdge} (a), (b) reports the parameter $\eta_n$, defined by Eq.~\eqref{eq:eta}, for these two configurations. In the \emph{dispersive} scenario, shown in Fig.~\ref{fig:CombEdge} (a), $\eta_n$ is largest for the modes nearest to $\omega_q$, namely $n=2$ and $n=3$, and decreases rapidly for more detuned modes. This behavior shows that the dynamics is governed primarily by a few nearby discrete modes rather than by a broad continuum of frequencies. Moreover, increasing the normalized reorganization rate $\lambda/\omega_q$ raises the overall magnitude of $\eta_n$ for all relevant modes, increasing the multimode character of the problem. The \emph{resonant} case, shown in Fig.~\ref{fig:CombEdge}(b), corresponds instead to $ \omega_q = \omega_3 $, for which the qubit is exactly resonant with the third TL mode. In this case, $\eta_3$ formally diverges because the detuning in the denominator vanishes. This divergence has no direct physical meaning: it simply indicates that $\eta_n$ is a useful figure of merit only for off-resonant modes, whereas at resonance the corresponding mode must be treated non-perturbatively and included explicitly in the system Hamiltonian. The remaining off-resonant modes still provide finite corrections, and their relative importance again increases with $\lambda/\omega_q$.

We now identify the modes that govern the dynamics in the \emph{dispersive} regime by solving the quantum dynamics with a varying finite number of TL modes, as shown in Fig.~\ref{fig:CombEdge}(c--e).  For the largest reorganization rate considered, $\lambda/\omega_q = 0.10$, the dynamics is highly sensitive to the number of TL modes retained in the truncated expansion. In particular, the single-mode-per-parity-sector approximation, in which only the TL modes with $n=2,3$ are included, already fails   at short times and predicts pronounced oscillations that are not representative of the converged multimode behavior. By contrast, the results obtained by retaining modes $n=1,\dots,10$, $n=1,\dots,20$, and $n=1,\dots,30$ become progressively closer to one another. This convergence indicates that, in the strong-coupling dispersive regime, a quantitatively reliable description requires the explicit inclusion of several off-resonant TL modes. As the reorganization rate is reduced, the convergence is substantially faster. For $\lambda/\omega_q = 0.05$, the curves obtained with $n=1,\ldots,20$, and $n=1,\ldots,30$ are nearly indistinguishable on the scale of the figure, while the single-mode-per-parity-sector truncation still exhibits clear deviations. For the weakest coupling considered, $\lambda/\omega_q = 0.025$, all truncations collapse almost completely onto the same slowly varying dynamics.
We also compare the multimode simulations with  corresponding dynamics obtained from the effective Hamiltonian of Eq.~\eqref{eq:EffectiveDispersiveHamiltonianSymmetric} (dashed line). The agreement is very good on the time scales shown in Fig.~\ref{fig:CombEdge}(c)--(e), with only small deviations appearing at long times, as expected from the approximations underlying the effective dispersive Hamiltonian in Eq.~\eqref{eq:EffectiveDispersiveHamiltonianSymmetric}. This trend is  consistent with the behavior of $\eta_n$ in Fig.~\ref{fig:CombEdge} ~(a): reducing $\lambda/\omega_q$ suppresses the effective hybridization with the nearby off-resonant modes, thereby weakening the multimode character of the comb-edge dynamics and making low-mode truncations progressively more accurate.

Figure~\ref{fig:CombEdge} (f--h) reports the analogous convergence analysis in the \emph{resonant} regime. In this case, the dynamics is governed primarily by the TL mode that is resonant with the qubit, which sets the dominant oscillation frequency.  However, the figure also shows that the remaining off-resonant modes cannot be neglected if quantitative accuracy is sought: they dress the resonant qubit--mode exchange and generate visible corrections to the oscillation amplitudes, phases, and revival pattern. This effect is most evident at the largest reorganization rate, $\lambda/\omega_q = 0.10$, shown in Figure~\ref{fig:CombEdge} (f), where the curves obtained with different truncations deviate appreciably over the time window, especially at intermediate and long times. In particular, while the lowest truncation captures the overall oscillatory character of the dynamics, it does not reproduce the detailed modulation induced by the multimode environment. Increasing the number of modes improves the agreement, but a quantitatively converged description requires the inclusion of a large number of TL modes. At $\lambda/\omega_q = 0.05$, the differences between successive truncations are already much smaller, indicating that the dressing from off-resonant modes is weaker. Finally, for $\lambda/\omega_q = 0.025$, all curves nearly collapse onto one another, showing that in the weak-coupling resonant regime the dynamics is effectively dominated by the single TL mode resonant with the qubits, with only minor corrections from the rest of the discrete spectrum.

\begin{figure}
\centering
\includegraphics[width=\linewidth]{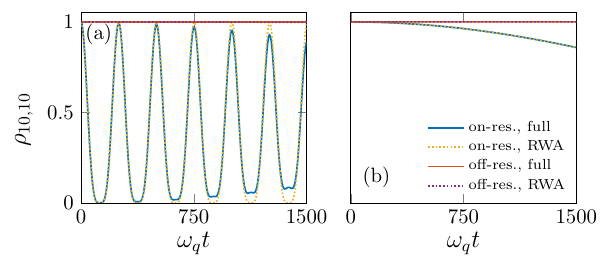}
\caption{Short-line single-mode region: Population dynamics of $\rho_{10,10}(t)$ for two qubits coupled through a finite transmission line with $\omega_g/\omega_{\mathrm{TL}}=0.1$. The initial state is $\ket{10}$, the coupling prefactor is fixed to $G=3.8\times 10^{-2}$, and the temperature is $k_B T/\hbar\omega_q=0.08$. Solid blue curves show resonant configurations, in which the qubits are tuned to a transmission line mode, while red curves show the corresponding off-resonant configurations, in which the qubit frequency lies midway between two neighboring modes. The yellow and purple dashed lines show the on-resonance and off-resonance configurations simulated within the RWA.
(a) Low-frequency realization: the resonant case has $\omega_q=\omega_1$, while the off-resonant case has $\omega_q=(\omega_1+\omega_2)/2$; the retained TL modes are $n=1,2$. 
(b) Higher-frequency realization: the resonant case has $\omega_q=\omega_5$, while the off-resonant case has $\omega_q=(\omega_4+\omega_5)/2$; the retained TL modes are $n=4,5$.}
\label{fig:SingleMode}
\end{figure}

\subsection{Short-line single-mode region}
\label{sec:ShortRegime}

Figure~\ref{fig:SingleMode} illustrates the dynamics in the short-line single-mode region (see Sec.~\ref{sec:ShortLineRegime}) for a short TL with  $\omega_g/\omega_\mathrm{TL} = 0.1$. The results are compared with simulations of the same system after applying the rotating-wave approximation (RWA). The populations $\rho_{10,10}(t)$ shown in Fig.~\ref{fig:SingleMode} are representative of two realizations of this operating region, corresponding to qubits operated at low (a) and higher frequencies (b).  In the single-mode region, the dynamics is controlled by the isolated TL mode closest to the qubit frequency. On-resonance, this mode dominates the coherent exchange, while the neighboring mode remains off-resonance. In addition, if the qubits are detuned from all modes, the dynamics is strongly suppressed. In both panels, the blue curve shows the resonant configuration, while the red curve shows the corresponding neighboring dispersive configuration.

In Fig.~\ref{fig:SingleMode}(a), the qubits are tuned either to the first TL mode, $\omega_q=\omega_1$, or midway between the first two modes. In the resonant case, $\rho_{10,10}(t)$ displays pronounced coherent oscillations, signaling excitation exchange with an isolated electromagnetic mode. The discrepancy from the RWA model that emerges at long times reflects the counter-rotating
terms retained in the full Hamiltonian.  In the off-resonant case, the dynamics is dispersive, and the population remains nearly constant over the investigated timescale because no mode is close enough in frequency to couple efficiently with the qubits. Figure~\ref{fig:SingleMode}(b) shows the analogous behavior near the fifth mode. The resonant dynamics is again effectively single-mode, but the oscillations are slower than in panel~(a), reflecting the decrease in coupling strength with increasing mode index. On this timescale, the counter-rotating terms leave no visible signature, and the full and RWA curves overlap.  The corresponding off-resonant configuration again yields an almost frozen population. This is consistent with the scaling $g_n \propto {1}/{\sqrt{n}}$ for large $n$ derived above: as the resonant mode index increases, the vacuum-Rabi frequency decreases, and the coherent population transfer correspondingly slows down. 


\section{Conclusions}
\label{sec:Conclusions}
 Transmission lines in superconducting circuits provide a realistic and controllable platform for engineering memory effects, collective dissipation, and coherent few-mode dynamics within the cQED picture. In this paper, we introduce a unified classification of the dynamical behavior of finite-length transmission lines in circuit quantum electrodynamics, which clarifies when the transmission line can behave as a structured reservoir, a discrete multimode coupler, or an effective single-mode resonator.

We considered a transmission line of length \(d\), characteristic impedance \(Z_{\mathrm{TL}}\), and propagation velocity \(v_p\) capacitively coupled to two transmon qubits with identical shunt capacitance \(C_J\) and Josephson energy \(E_J\), via two equal coupling capacitances \(C_c\). Starting from first-principles circuit quantization, we derived a Hamiltonian in which the transmission line modes separate naturally into even- and odd-parity sectors, coupled to the collective qubit operators $\hat L_\pm = \hat\sigma_y^{(1)} \pm \hat\sigma_y^{(2)}$. This decomposition provides a unified framework to analyze how the reduced dynamics is controlled by the hierarchy among the qubit frequency $\omega_q$, the transmission line characteristic frequency $\omega_{\mathrm{TL}} = 2 \pi {v_p}/{d} $, and the characteristic frequency $\omega_g = 2 \pi / (Z_{\mathrm{TL}} C_s)$ where $C_s = C_JC_c/ (C_J+ C_c)$.

In the long-line continuum region, $\omega_{\mathrm{TL}} \ll \omega_g,\omega_q$, each parity sector can be described as an independent bosonic reservoir with a Drude--Lorentz spectral density whose parameters are determined directly by the circuit elements. Using the hierarchical equations of motion (HEOM), we characterized the qubit reduced dynamics beyond perturbative and Markovian approximations and quantified memory effects through the Breuer--Laine--Piilo (BLP) measure of non-Markovianity. We found that non-Markovian effects are strongest for slow baths, while increasing the bath relaxation rate drives the dynamics toward an effectively Markovian regime. By comparing HEOM with the secular GKLS and TCL2 master equations, we showed that weak information backflow does not necessarily imply quantitative agreement with perturbative weak-memory treatments.

When either of the long-line conditions is violated, the continuum 
approximation breaks down, and the qubit reduced dynamics is governed by a discrete  set of line modes. We refer to this complementary region of parameter space as the \textit{discrete-mode region} and identify two limiting cases that admit an especially simple description. In the \textit{comb-edge discrete region} 
($\omega_{\mathrm{TL}} \ll \omega_g$, $\omega_q \lesssim \omega_{\mathrm{TL}}$), the qubit frequency lies close to the low-frequency edge of the spectrum, and  the behavior is intrinsically multimode, depending sensitively on whether the qubit is tuned resonantly or dispersively with respect to the nearby line  modes. This region is particularly important for superconducting-circuit 
applications, since it corresponds to the common operating point in which a  transmission line is used as a coupler between qubits and is frequently, yet too naively, approximated as an effective single-mode resonator. In the  \textit{short-line region} ($\omega_g \ll \omega_{\mathrm{TL}}$), the line  spectrum is sparse on the scale of the qubit--line hybridization and, for typical parameters of superconducting quantum devices, the dynamics is dominated by a single isolated mode, giving rise to cavity-like coherent  exchange rather than reservoir-induced relaxation. Away from these two limits, the number of modes required for a converged description depends nontrivially 
on the hierarchy among $\omega_q$, $\omega_g$, and $\omega_{\mathrm{TL}}$, and 
on the value of $G$; it can, in general, be large.

The present framework can be extended in several directions, including qubits coupled asymmetrically to the transmission line.  In the long-line continuum region, this has two important implications. First, the density of states is no longer approximately homogeneous over the relevant frequency window. Second, the two qubits couple to distinct effective environments, characterized by different spectral densities. These features are expected to further enrich the reduced dynamics, leading to an interplay of asymmetry and structured dissipation that deserves dedicated investigation.

Taken together, these results provide a unified cQED picture of how a finite superconducting transmission line interpolates between three physically distinct roles: a structured environment, a few-mode quantum coupler, and an effectively single-mode quantum resonator. More broadly, the present work contributes a first-principles regime map that connects circuit parameters, effective bath structure, and the validity of Markovian, multimode, and single-mode descriptions within a single cQED model. This establishes explicit criteria for the dynamical regimes relevant to present and future superconducting-circuit experiments.

\begin{acknowledgements}
We are grateful to I\~nigo Luis Egusquiza from the University of the Basque Country UPV/EHU for useful suggestions and discussions. This work was supported by the Italian Ministry of University and Research through the PNRR MUR Project No. PE0000023-NQSTI. A. P.-R. acknowledges support from the European Union’s Marie Skłodowska-Curie Actions under grant agreement No. 101204967 (FTMcQED).
\end{acknowledgements}

\appendix
\section{Normal Mode Expansion}
\label{app:Normalization}

The transmission line (TL) modes $u_n(x)$ of the scenario displayed in Fig.~\ref{fig:finiteTLoverview} (a) of a TL of length $d$, symmetrically coupled to two transmon qubits, are the solutions of the following auxiliary eigenvalue problem introduced in Ref. \cite{parra-rodriguez_quantum_2018}:
\begin{equation}
u_n^{\prime \prime}(x) =-k_n^2 u_n(x), \qquad \text{with} \quad  0 < x < d
\label{eq:EquationEig}
\end{equation}
with the following boundary conditions at the two line ends $x=0$ and $x=d$:
\begin{subequations}
\begin{align}
\label{eq:BC_o}
u_n^{\prime}(0)&=-k_n^2 \alpha \, u_n(0), \\ 
\label{eq:BC_d}
u_n^{\prime}(d)&=+k_n^2 \alpha \, u_n(d),
\end{align}
\label{eq:BCs}
\end{subequations}
where $\alpha$ is a parameter  fixed  in order to remove TL mode-mode coupling.
The general solution of Eqs. \eqref{eq:EquationEig} and \eqref{eq:BCs} is
\begin{equation}
u_n(x)=
\begin{cases}
\displaystyle
+A_n\,
\frac{\cos\!\left[k_n\left(x-{d}/{2}\right)\right]}{\cos\!\left[k_n {d}/{2} \right]},
& n=0,2,4,\ldots,
\\[10pt]
\displaystyle
-\,A_n\,
\frac{\sin\!\left[k_n\left(x-{d}/{2}\right)\right]}{\sin\!\left[k_n {d}/{2} \right]},
& n=1,3,\ldots,
\end{cases}
\label{eq:un}
\end{equation}
where $A_n=|u_n(0)|=|u_n(d)|$. By  combining Eq.~\eqref{eq:un} and its derivative in Eq.~\eqref{eq:BCs} we obtain the two characteristic equations \eqref{eq:even_eig} and  \eqref{eq:odd_eig} for even and odd modes.

Since the eigenvalue appears in the boundary conditions \eqref{eq:BCs}, the modes $\{ u_n \}$ are orthogonal with respect to the augmented scalar product:
\begin{align}
    \label{eq:Norm_finite}
\begin{aligned}
    &\int_0^d  u_n(x)u_m(x) dx  +\alpha u_n(0)u_m(0) + \alpha u_n(d)u_m(d)  \\  
   &=  \frac{N_{\alpha}}{c} \delta_{nm},
\end{aligned}
\end{align}
where $N_\alpha$ is a normalization factor. 
Substituting expressions \eqref{eq:un} into Eq.~\eqref{eq:Norm_finite} with \(n=m\) we obtain:
\begin{equation}
A_n= 
\begin{cases}
    \sqrt{\dfrac{N_\alpha}{c(d+2\alpha)}} & n=0; \\
    \sqrt{ \dfrac{N_\alpha} {c\left[ {d}/{2}\left(1+\alpha^2  k_n^2\right)+\alpha
\right]}}  & n=1,2,\ldots.
\end{cases}
\label{eq:AnO}
\end{equation}

The scalar product \eqref{eq:Norm_finite} becomes an ordinary Hilbert-space inner product when the
endpoint amplitudes are treated as additional components of an enlarged
space. Related generalized endpoint sum rules also
follow from the doubled-space formulation of
Ref.~\cite{parra-rodriguez_exact_2025}.
The completeness of the eigenmodes in the augmented space implies the endpoint sum rules
\begin{equation}
\uvec_0^{\intercal}\uvec_d=0,
\qquad
\uvec_0^{\intercal}\uvec_0
=
\uvec_d^{\intercal}\uvec_d
=
\frac{N_\alpha}{c\alpha},
\label{eq:endpoint_sum_rules}
\end{equation}
where 
$\mathbf{u}_0 = \left(u_0(0), u_1(0),...,u_n(0),\ldots \right)^T$ and
$\mathbf{u}_d = \left(u_0(d), u_1(d),...,u_n(d),\ldots \right)^T$. In the paper, we have adopted the normalization
$\mathbf{u}_0^\intercal\mathbf{u}_0=\mathbf{u}_d^\intercal\mathbf{u}_d=1$, which is equivalent to fixing
$N_\alpha=c\alpha$.

\section{Hamiltonian}
\label{app:ClassicalHam}
We start from the circuit Lagrangian and expand the transmission line flux field and charge density in terms of the normal modes introduced in the previous section. As shown in Ref.~\cite{parra-rodriguez_quantum_2018}, the choice
\begin{equation}
  \alpha = {C_s}/{c}  
\end{equation}
diagonalizes the transmission line  Hamiltonian, thereby removing transmission line mode--mode couplings. From the adopted normalization for the modes, it follows that
$N_\alpha={C_s}$.

Then, the circuit Hamiltonian can be written as
\begin{equation}
    \mathcal{H}
    =
    \mathcal{H}_{t1}
    +
    \mathcal{H}_{t2}
    +
    \mathcal{H}_{\mathrm{TL}}
    +
    \mathcal{H}_{\mathrm{int}} .
    \label{eq:Hamiltonian_first_quantized_complete_basis}
\end{equation}
The qubit transmon Hamiltonians are
\begin{equation}
    \mathcal{H}_{ti}
    =
    \frac{Q_i^2}{2C_J}
    -
    E_J
    \cos\left(
        \frac{2\pi \Phi_i}{\Phi_Q}
    \right),
    \qquad i=1,2 .
\end{equation}
The transmission line Hamiltonian, written in the complete auxiliary-mode basis, is
\begin{equation}
    \mathcal{H}_{\mathrm{TL}}
    =
    \sum_{n=0}^{\infty}
    \left[
        \frac{q_n^2}{2C_s}
        +
        \frac{1}{2}C_s \omega_n^2 \varphi_n^2
    \right] .
\end{equation}
The interaction Hamiltonian reads
\begin{equation}
    \mathcal{H}_{\mathrm{int}}
    =
    \frac{Q_1}{C_J}
    \sum_{n=0}^{\infty} u_n(0) q_n
    +
    \frac{Q_2}{C_J}
    \sum_{n=0}^{\infty} u_n(d) q_n .
\end{equation}
In the Hamiltonian of Eq.~\eqref{eq:Hamiltonian_first_quantized_complete_basis},
no direct qubit--qubit coupling term appears. This cancellation follows from
the completeness of the transmission line eigenmodes with respect to the
augmented inner product, as expressed by the endpoint sum rules in
Eq.~\eqref{eq:endpoint_sum_rules}. 

The zero-frequency mode must be excluded from the quantized transmission line
oscillator expansion. Indeed, the $n=0$ even mode has $\omega_0=0$ and $u_0(x)=A_0$, with
$A_0^2=N_\alpha/[c(d+2\alpha)]$. Since $\varphi_0$ does not appear in
the Hamiltonian, its conjugate momentum $q_0$ is conserved in time.  The evolution of \(\varphi_0\) is therefore fixed by the conserved value of \(q_0\), together with the qubit charges \(Q_1(t)\) and \(Q_2(t)\), whose dynamics is independent of \(\varphi_0\).  Once the physical oscillator modes and the qubit variables have been specified, $(\varphi_0,q_0)$ is fixed and should not be included in the quantization.
 From a physical point of view, the zero mode is a global electrostatic degree of freedom associated with a uniform shift of the
node charges, which does not contribute to the dynamics of the system. This offset can be absorbed through a gauge
transformation of the superconducting phases.

We finally promote the dynamical flux and charge variables with \(n\geq 1\) to operators satisfying
\begin{equation}
    \left[\hat{\varphi}_n,\hat{q}_m\right]
    =
    i\hbar \delta_{nm},
    \qquad n,m\geq 1 .
\end{equation}
This gives the quantized Hamiltonian in Eq.~\eqref{eq:QuantizedHamiltonian}.

\section{Experimental context of the regime map}
\label{sec:LiteraturePlacement}

To assign a point on the parameter plane \((\omega_g/\omega_{\mathrm{TL}},\omega_q/\omega_{\mathrm{TL}})\) of Fig. 1(d) to the cited experiments, we proceed as follows. First, we identify the characteristic transmission line scale $\omega_{\mathrm{TL}}$. For a conventional $\lambda/2$ resonator, we use $\omega_{\mathrm{TL}}=2\omega_r$ \cite{dicarlo_demonstration_2009, filipp_multimode_2011, filipp_two-qubit_2009, houck_controlling_2008, majer_coupling_2007}, while for a $\lambda/4$ resonator, we use $\omega_{\mathrm{TL}}=4\omega_r$ \cite{dumur_v-shaped_2015}, where $\omega_r$ is the reported resonance angular frequency. For long cable or coplanar-waveguide interconnects supporting many standing modes, we use the reported free spectral range as $\omega_{\mathrm{TL}}$ \cite{qiu_thermal-noise-resilient_2026, qiu_deterministic_2025, heya_randomized_2025, song_realization_2025}. For devices that are not literal finite transmission lines, such as three-dimensional cavities \cite{rettaroli_novel_2025}, tunable lumped couplers \cite{xu_high-fidelity_2020}, or metamaterial ring resonators  \cite{mcbroom-carroll_entangling_2024}, we use the relevant bus-mode frequency or cutoff frequency as an effective frequency $\omega_{\mathrm{TL}}^{\mathrm{eff}}$. The vertical coordinate $ \omega_q/\omega_{\mathrm{TL}}$ is then obtained directly from the reported qubit frequency \cite{dicarlo_demonstration_2009, filipp_multimode_2011, mcbroom-carroll_entangling_2024, xu_high-fidelity_2020, qiu_thermal-noise-resilient_2026, qiu_deterministic_2025, heya_randomized_2025, song_realization_2025, majer_coupling_2007}.

When the qubit frequency is not explicitly given, it is estimated from the transmon expression $\frac{\omega_q}{2 \pi} \simeq \frac{1}{h}\sqrt{8E_JE_C}-\frac{E_C}{h}$ \cite{nongthombam_entanglement_2026, darrigo_optimal_2012}.  Whenever the relevant capacitances are reported, as in Ref. \cite{mcbroom-carroll_entangling_2024, rettaroli_novel_2025}, we compute $C_s$ directly.

If the capacitances are not directly available, but a qubit--mode coupling $g_n$ is reported, we infer an effective value of $C_s$ by inverting the single-mode coupling expression derived in our model \cite{dicarlo_demonstration_2009, filipp_two-qubit_2009, filipp_multimode_2011, houck_controlling_2008, xu_high-fidelity_2020, qiu_thermal-noise-resilient_2026, qiu_deterministic_2025, heya_randomized_2025, song_realization_2025, majer_coupling_2007, pita-vidal_direct_2023}. In this case, the extracted position should be interpreted as an effective circuit point rather than as a purely geometrical value. Since most references do not report \(Z_{\mathrm{TL}}\), we assume the standard cQED value $Z_{\mathrm{TL}}=50~\Omega$. Whenever possible, as in \cite{mcbroom-carroll_entangling_2024}, the reported capacitance and inductance per unit length are instead used to reconstruct the effective characteristic impedance.

The charge matrix element entering the coupling is reconstructed using the harmonic transmon approximation
 $n_T \simeq\left(
    \frac{E_J}{32E_C}
    \right)^{1/4}$\cite{dicarlo_demonstration_2009, filipp_two-qubit_2009, mcbroom-carroll_entangling_2024, song_realization_2025}.
When \(E_J\) and \(E_C\) are not both reported, we estimate \(E_C\) from the transmon anharmonicity $E_C/h$ and then infer \(E_J/E_C\) from the reported value of $\omega_q$ \cite{xu_high-fidelity_2020, rettaroli_novel_2025, qiu_thermal-noise-resilient_2026, qiu_deterministic_2025, heya_randomized_2025}.

At the Hamiltonian level, the references collected in Fig.~\ref{fig:finiteTLoverview}(d)
cover different effective descriptions. Conventional cavity-QED devices are
typically described by single-mode or few-mode Jaynes--Cummings/dispersive
Hamiltonians, where two transmons couple to a selected resonator mode \cite{fink_dressed_2009,dicarlo_demonstration_2009, filipp_two-qubit_2009, houck_controlling_2008, majer_coupling_2007}. Multimode cavity and
interconnect experiments instead require a sum over standing-wave modes of the
bus or resonator \cite{filipp_multimode_2011, qiu_thermal-noise-resilient_2026, heya_randomized_2025, song_realization_2025}. Other works are more naturally
formulated in input--output or cascaded-system language, with tunable emission
and absorption rates into a propagating microwave channel \cite{magnard_microwave_2020, xiang_intracity_2017}.

\section{Finite-temperature error of the continuum approximation}
\label{sec:TemperatureRole}
Since the lowest transmission line mode appears at frequency
\(\omega_{\mathrm{TL}}/2\) in the odd sector and at frequency
\(\omega_{\mathrm{TL}}\) in the even sector, the continuum approximation assigns spectral weight to a frequency interval where no discrete mode is actually present. This produces an error \(\varepsilon_C(\tau)\) in the bath correlation function, associated with the intervals \(\omega\in[0,\omega_{\mathrm{TL}}/2]\) for the odd sector and
\(\omega\in[0,\omega_{\mathrm{TL}}]\) for the even sector. We evaluate
\(\varepsilon_C(\tau)\) using the odd-sector cutoff:

\begin{multline}
\label{eq:Cerror}
\frac{\varepsilon_C(\tau)}{\hbar^2} \\ =  \int_0^{\omegaTL/2}\!d\omega\;\frac{J(\omega)}{\pi}
\left[\coth\!\left(\frac{\beta\hbar\omega}{2}\right)\cos(\omega\tau)-i\sin(\omega\tau)\right].
\end{multline}
Temperature enters only through the real part of this error
because the thermal factor multiplies only the cosine contribution:
\begin{equation}
\Re  \varepsilon_C(\tau) =\hbar^2\int_0^{\omegaTL/2} d\omega\,\frac{J(\omega)}{\pi}\coth\!\left(\frac{\beta\hbar\omega}{2}\right)\cos(\omega\tau),
\label{eq:ReC}
\end{equation}
whereas the imaginary part,
\begin{equation}
\Im \varepsilon_C(\tau)=-\hbar^2\int_0^{\omegaTL/2} d\omega\,\frac{J(\omega)}{\pi}\sin(\omega\tau),
\label{eq:ImC}
\end{equation}
is temperature independent.
In the regime  $\hbar \omegaTL \ll k_B T$, the 
high-temperature expansion
\begin{equation}
\coth\!\left(\frac{\beta\hbar\omega}{2}\right)\simeq  2 \frac{k_B T}{\hbar \omega},
\end{equation}
holds throughout the integration domain. Combined with the 
low frequency behavior of the Drude--Lorentz spectral density, 
$J(\omega) \simeq 2\lambda\omega/\gamma$ for $\omega \ll \gamma$ 
(which is guaranteed in the long-line continuum region, 
$\gamma \simeq \omega_g/(2\pi) \gg \omegaTL$), we obtain
\begin{equation}
\frac{J(\omega)}{\pi}\coth\!\left(\frac{\beta\hbar\omega}{2}\right)
\simeq
\frac{4\lambda}{\pi\gamma}\, \frac{k_B T}{\hbar}.
\end{equation}
For sufficiently short times $\omegaTL\tau\ll 1$, one may further approximate $\cos(\omega\tau)\simeq 1$, so that
\begin{equation}
\Re\,\varepsilon_C(\tau) \sim 
\frac{2\lambda}{\pi\gamma}\,k_B T\,\hbar\,\omegaTL.
\end{equation}
The error of the continuum approximation on $\Re C(\tau)$ thus
grows linearly with both the temperature and the mode spacing
$\omegaTL$. By evaluating the correlation function Eq.~\eqref{eq:Cerror} at zero temperature in the same low-frequency and short-time regime, one finds
\begin{equation}
\Re\varepsilon_C^{0}(\tau)
\simeq
\frac{\hbar^2\lambda}{4\pi\gamma}\omega_{\mathrm{TL}}^2 .
\end{equation}
Therefore, finite-temperature enhancement of the spurious low-frequency contribution with respect to the zero-temperature case is
\begin{equation}
\frac{\Re\varepsilon_C(\tau)}
{\Re\varepsilon_C^{0}(\tau)}
\simeq
8\frac{k_B T}{\hbar\omega_{\mathrm{TL}}}.
\label{eq:Cerror_ratio}
\end{equation}
To estimate the relative weight of this error within the full continuum correlation function, we introduce the dimensionless ratio
\begin{equation}
\delta_C(\tau)
=
\frac{\left|\Re\varepsilon_C(\tau)\right|}
{\left|\Re C(\tau)\right|}.
\end{equation}
For the representative cases shown in Fig.~\ref{fig:CorrelationFunctions}, we evaluate this ratio at $\tau=0$, where the real part of the low-frequency error reaches its maximum value, and report the corresponding values in Table~\ref{tab:C_error_low}.

\begin{table}[t]
  \centering
\small
  \setlength{\tabcolsep}{6pt}
  \renewcommand{\arraystretch}{1.15}
  \caption{Relative size of the odd-sector low-frequency continuum error at $T=0$ and $\tau=0$.}
  \label{tab:C_error_low}
  \begin{tabular}{c c}
    \toprule
    $\omega_{\mathrm{TL}}/\omega_g$
    &
    $\delta_C(0)$ \\
    \midrule
    $10$  & $1.6\times 10^{-1}$ \\
    $1$   & $6.2\times 10^{-2}$ \\
    $0.1$ & $4.7\times 10^{-3}$ \\
    \bottomrule
  \end{tabular}
\end{table}

The most relevant case for the continuum approximation is $\omega_{\mathrm{TL}}/\omega_g=0.1$, for which the relative error is only $4.7\times 10^{-3}$. This confirms that, in the long-line continuum regime, the spurious low-frequency contribution is negligible compared with the full continuum correlation function, and it is also consistent with the usual working temperature of superconducting systems.

\section{Order-of-magnitude estimates}
\label{sec:OrderMagnitude}

To identify the physically relevant parameter range explored in the numerical
analysis, we provide order-of-magnitude estimates based on typical
superconducting-circuit values reported in the recent
literature~\cite{krantz_quantum_2019,blais_circuit_2021,bosman_approaching_2017,fink_dressed_2009,chen_transmon_2023}. For a transmon operating in the standard microwave-frequency
window~\cite{krantz_quantum_2019,blais_circuit_2021}, we take
\begin{equation}
\omega_q = 2\pi \times 5~\mathrm{GHz},
\qquad
C_J= 50\text{--}200~\mathrm{fF},
\end{equation}
where the shunt-capacitance range covers typical planar transmon
designs~\cite{krantz_quantum_2019,place_new_2021}.

For the transmission line, an impedance-matched coplanar waveguide
has
\begin{equation}
Z_{\mathrm{TL}} \simeq 50~\Omega,
\qquad
v_p \simeq 0.8\text{--}1.5 \times10^{8}~\mathrm{m/s},
\end{equation}
as in standard circuit-QED implementations~\cite{blais_circuit_2021,besedin_quality_2018}.
However, in order to connect the numerical scan to both standard and
high-impedance superconducting circuits, we regard the product
$Z_{\mathrm{TL}}C_s$ as the relevant parameter controlling
the order of magnitude of the bath relaxation rates. We consider as effective range
\begin{equation}
Z_{\mathrm{TL}} \simeq 50~\Omega\text{--}1.6~\mathrm{k}\Omega,
\quad
C_s \simeq 1\text{--}100~\mathrm{fF},
\end{equation}
where the upper part of the impedance range should be understood as
representative of high-impedance resonators or strongly loaded transmission line
environments~\cite{bosman_approaching_2017}. 
Consequently, the propagation velocity results in an effective range 
\begin{equation}
    v_p \simeq 10^7\text{--}1.5\times 10^8~\mathrm{m/s}.
\end{equation}
We further assume a Cooper-pair
transition matrix element $n_T \approx 1$, consistent with the transmon
regime~\cite{krantz_quantum_2019}. We obtain
\begin{equation}
\tau_g \equiv Z_{\mathrm{TL}} C_s
\simeq 0.05\text{--}160~\mathrm{ps},
\end{equation}
and hence the value of the characteristic coupling frequency is in the range \eqref{eq:omegag} 
\begin{equation}
\frac{\omega_g}{2\pi} = \frac{1}{\tau_g}
\simeq 6\text{--}2\times 10^{4}~\mathrm{GHz}.
\end{equation}
Using Eq.~\eqref{eq:gamma} and the long-line limit
$\gamma \simeq \omega_g/(2\pi)$, we obtain
\begin{equation}
\frac{\gamma}{\omega_q}
\simeq 0.2\text{--}6\times 10^{2}.
\end{equation}

For the reorganization rate, the estimates obtained from the expression
\eqref{eq:lambda} give the order of magnitude range
\begin{equation}
\frac{\lambda}{\omega_q}
\simeq 10^{-4}\text{--}10^{-1}.
\end{equation}

Finally, superconducting qubits are typically operated at the base stage of a
dilution refrigerator, i.e. at $T \simeq 10\text{--}20~\mathrm{mK}$~\cite{krantz_quantum_2019,blais_circuit_2021}.
In the normalized units adopted throughout this work,
$T = 20~\mathrm{mK}$ corresponds to
\begin{equation}
\frac{\kb T}{\hbar \omega_q} \simeq 0.08.
\end{equation}

\section{Qubit coupled to a short-circuited transmission line}
\label{sec:LineShort}

    \begin{figure}
        \centering
        \includegraphics[width=\linewidth]{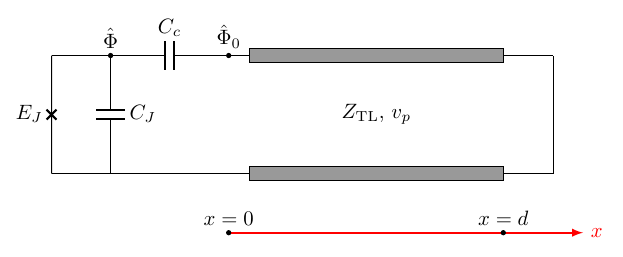}
        \caption{Schematic of a short-circuited transmission line capacitively coupled to a transmon qubit.}
        \label{fig:LineShort}
    \end{figure}

Here, we consider the circuit shown in Fig.~\ref{fig:LineShort}, consisting of a short-circuited transmission line (TL) of length $d$ capacitively coupled to a transmon qubit. The transmission line is characterized by the characteristic impedance \(Z_{TL}\) and propagation velocity $v_p$. 

By applying the same approach used for the TL capacitively coupled to two transmon qubits, we obtain the Hamiltonian
\begin{equation}
\hat{H}=\hat{H}_{t}+\hat{H}_{\mathrm{TL}}+\hat{H}_{\mathrm{int}}
\end{equation}
where
\begin{equation}
\hat{H}_{t}\approx \frac{\hbar\omega_q}{2}\,\hat{\sigma}_z,
\label{eq:Hsys_1Q}
\end{equation}
\begin{equation}
\hat H_{\mathrm{TL}}
=\sum_{n=1}^\infty\hbar\omega_n\left(\hat b_n^\dagger\hat b_n+\frac{1}{2}\right),
\end{equation}
and
\begin{equation}\label{eq:Hint_bosons_1q}
\hat{H}_{\mathrm{int}}=\hbar \hat{\sigma}_y 
\sum_{n=1}^\infty g_n \left(\hat{b}_n+\hat{b}_n^{\dagger}\right).
\end{equation}
The mode frequencies $\{\omega_n\}$ are given by $\omega_n=v_pk_n$ where the wavenumbers $\{k_n\}$ are solutions of the transcendental equation
\begin{equation}
\label{eq:eig_1Q}
    \cot(k_nd) = \frac{C_s}{c} k_n.
\end{equation}
The mode-dependent coupling strength $g_n$ is given by
\begin{equation}\label{eq:gn_1Q}
g_n
= G \sqrt{\frac{\omega_\mathrm{TL}\omega_n}{ \left( 2 \pi \omega_n / \omega_g\right)^2 + \omegaTL / \omega_g + 1 }},
\end{equation}
where the characteristic frequency $\omega_g$, the transmission line characteristic frequency $\omegaTL$ and the dimensionless factor $G$ are defined in Eqs. \eqref{eq:omegag}, \eqref{eq:omegaTL} and \eqref{eq:FactorG}, respectively.

Equation~\eqref{eq:eig_1Q} can be solved numerically. In the parameter region typical of circuit QED here, ${C_s}/{c} \ll 1$ so that to an excellent approximation one may neglect the right hand side
\begin{equation}
k_n d  \simeq  \frac{(2n-1)\pi}{2},\qquad n=1,2,\ldots,
\end{equation}
and the corresponding mode frequencies are
\begin{equation}
\omega_n  \simeq 
v_p\,\frac{(2n-1)\pi}{2d}= \frac{(2n - 1)}{4}\,\omega_\text{TL}.
\label{eq:FrequencySpacing_1Q}
\end{equation}

As already discussed in Sec.~\ref{sec:DynamicalRegimes}, the transmission line can be treated as a continuum when both of the conditions in Eq.~\eqref{eq:LongLineConditions} hold. In this limit, the line is characterized by the same spectral density given in Eq.~\eqref{eq:Jomega_DL} with 
\begin{subequations}
\begin{align}
\label{eq:GammaLong_1Q}
\gammaONEq  & \approx  \omega_g / (2\pi), \\
\lambdaONEq & \approx \pi G^2 \gammaONEq.
\label{eq:LambdaLong_1Q}
\end{align}
\end{subequations}

\section{Hierarchical equations of motion and simulation details}

\label{sec:HEOMCorrelationExpansion}

The hierarchical equations of motion (HEOM), originally introduced by Tanimura and Kubo~\cite{tanimura_time_1989}, provide a non-perturbative formulation of the reduced dynamics of a quantum system linearly coupled to a Gaussian bosonic environment. The method rewrites the non-local memory of a discrete environment or a reservoir encoded in the correlation function as a hierarchy of coupled time-local equations for the reduced density operator and a set of auxiliary density operators (ADOs). The reduced density operator corresponds to the zeroth-tier element of the hierarchy, while the higher-tier ADOs encode progressively higher orders of system--environment memory.

The HEOM construction requires the correlation functions $C(t)$ to be represented, exactly or approximately, as finite sums of exponentials. In fact, by separating the real and imaginary parts, we write
\begin{equation}
C(t)=C_R(t)+i C_I(t),
\end{equation}
where $C_R(t)$ and $C_I(t)$ are expanded as
\begin{equation}
C_R(t)/\hbar^2  =
\sum_{k=1}^{N_R}
c_k^R e^{-\gamma_k^R t}, \quad 
C_I(t)/\hbar^2 =
\sum_{k=1}^{N_I}
c_k^I e^{-\gamma_k^I t}.
\label{eq:HEOMCorrelationExpansion}
\end{equation}
The coefficients \(c_k^{R,I}\) and rates \(\gamma_k^{R,I}\) may, in general, be complex, provided that the full sums in Eq.~\eqref{eq:HEOMCorrelationExpansion} reproduce the real-valued functions \(C_R(t)\) and \(C_I(t)\). This exponential representation allows one to introduce a finite set of auxiliary density operators and to obtain a closed hierarchy of first-order differential equations:

\begin{widetext}

\begin{align}
\begin{aligned}
\dot{\rho}^{\mathbf{n}}(t) =&\left(\mathcal{L}-\sum_{\alpha=\pm} \sum_{j=R, I} \sum_{k=1}^{N_j} n_{\alpha j k} \gamma_k^j\right) \rho^{\mathbf{n}}(t)
 -i \sum_{\alpha=\pm} \sum_{k=1}^{N_R} c_k^R n_{\alpha R k} \hat{L}_\alpha^{\times} \rho^{\mathbf{n}_{\alpha R k}^{\downarrow}}(t)+\sum_{\alpha=\pm} \sum_{k=1}^{N_I} c_k^I n_{\alpha I k} \hat{L}_\alpha^{\circ} \rho^{\mathbf{n}_{\alpha I k}^{\downarrow}}(t) \\
 &-i \sum_{\alpha=\pm} \sum_{j=R, I} \sum_{k=1}^{N_j} \hat{L}_\alpha^{\times} \rho^{\mathbf{n}_{\alpha j k}^{\uparrow}}(t)
 \end{aligned}
 \label{eq:HEOMeq}
\end{align}
\end{widetext}
where $\hat{L}^{\times}\bullet=[\hat{L},  \bullet  ]$, $\quad \hat{L}^{\circ}\bullet=\{\hat{L},  \bullet  \}$,
$\mathcal L\bullet=-(i/\hbar)[\hat H_S,\bullet]$ and $\hat H_S$ is the system Hamiltonian. The label $\mathbf{n}=(\mathbf{n}_{+},\mathbf{n}_{-})$, with
$\mathbf{n}_{\alpha}=\left(n_{\alpha R 1},\ldots,n_{\alpha R N_R},
n_{\alpha I 1},\ldots,n_{\alpha I N_I}\right)$, is a multi-index of
non-negative integers subject to the tier constraint
$\sum_{\alpha,j,k} n_{\alpha j k}\le N_c$, where $N_c$ is the hierarchy
depth, a cutoff parameter chosen for convergence. 
The notation $\rho^{\mathbf n_{\alpha j k}^{\uparrow}}$ and
$\rho^{\mathbf n_{\alpha j k}^{\downarrow}}$ denotes the ADO obtained from
$\rho^{\mathbf n}$ by increasing or decreasing the component
$n_{\alpha j k}$ by one. The density operator labeled by $\mathbf{n}=\left(0,\ldots,0\right)$ is the reduced density operator of the system, while the elements with nonzero indices are the ADOs.

In the limit in which both the exponential representation and the hierarchy depth are converged, the HEOM dynamics is numerically exact for
the corresponding linear bosonic environment. For a detailed derivation of Eq.~\eqref{eq:HEOMeq} and implementation details, we refer the reader to Refs.~\cite{tanimura_numerically_2020,lambert_qutip-bofin_2023}.
The HEOM calculations were performed in Python using the QuTiP--BoFiN package developed by Lambert \textit{et al.}~\cite{lambert_qutip-bofin_2023}. Since each parity sector contributes \(N_R+N_I\) exponential terms, the multi-index \(\mathbf{n}\) has \(K=2(N_R+N_I)\) components, and truncation at tier \(N_c\) yields a total number of ADOs~\cite{tanimura_numerically_2020}
\begin{equation}
N_{\mathrm{ADO}}
=
\sum_{\ell=0}^{N_c}
\binom{K+\ell-1}{\ell}
=
\binom{K+N_c}{N_c}.
\end{equation}
The computational cost therefore grows rapidly with both the number of exponential terms retained in the correlation function decomposition and the hierarchy depth.
In the results described in Sec.~\ref{sec:Results}, the system Hamiltonian used to compute the Liouvillian is the one in Eq.~\eqref{eq:Hsys} and Eq.~\eqref{eq:Hsys_1Q}.

In the long-line continuum region, the two parity sectors have identical correlation functions, \(C_{+}(t)=C_{-}(t) \approx C(t)\), with \(C(t)\) given by Eq.~\eqref{eq:CorrFunct_Cont}. In this case, at thermal equilibrium, the bath correlation function was first generated from a Matsubara expansion. This representation was then compressed by a least-squares fitting procedure. Unlike the real part, the imaginary part of the Drude--Lorentz correlation function is given analytically by a single exponential and contains no temperature-dependent contribution, so \(N_I=1\) throughout. In the low-temperature and large bath relaxation rate, we used \(N_R=3\), corresponding to \(K=8\), while \(N_R=4\) (\(K=10\)) was required to maintain convergence for small bath relaxation rates. We verified in both cases that the solutions do not significantly change by increasing \(N_R\) to 5.

For the BLP non-Markovianity calculations, convergence was typically reached already at hierarchy depth \(N_c=3\), while a small subset of low-temperature parameter points required \(N_c=4\). We verified that the solutions do not significantly change by increasing $N_c$ to 5. In addition, in order to resolve the trace-distance dynamics accurately, the simulations were propagated up to a final time
$ \omega_q t_{\mathrm{end}} = 100$, even when the relevant populations had already approached their stationary values at earlier times.

In the discrete-mode region, the correlation function was constructed directly from the discrete transmission line spectrum, namely from the mode frequencies \(\omega_n\) and couplings \(g_n\), without any compression step. Convergence was obtained with \(N_c=3\). However, because no compression of the correlation function was performed, the number \( K\) of exponential terms entering the HEOM construction can become very large, especially when many transmission line modes must be retained. As a consequence, the numerical cost in the discrete-mode region can be substantially higher than in the continuum case, even when the hierarchy depth is the same.

\section{BLP non-Markovianity measure}
\label{sec:BLP}
The flow of information between an open quantum system and its environment can be tracked over time through the evolution of the trace distance between two reduced density operators of the system, obtained starting from different initial conditions. Since the trace distance quantifies the distinguishability of quantum states, its evolution provides an operational way to assess whether information is being irreversibly lost to the environment or temporarily recovered by the system.

The trace norm of an operator $A$ is defined by \cite{breuer_theory_2002}
\begin{equation}
\|A\|=\operatorname{tr}|A|=\operatorname{tr}\sqrt{A^{\dagger}A}.
\end{equation}
If $A$ is self-adjoint with eigenvalues $a_i$, then $\|A\|=\sum_i |a_i|$ (counting multiplicities). For two density operators $\rho^1$ and $\rho^2$, the induced trace distance is
\begin{equation}
D(\rho^1,\rho^2)
=\frac{1}{2}\|\rho^1-\rho^2\|
=\frac{1}{2}\operatorname{tr}\bigl|\rho^1-\rho^2\bigr|.
\end{equation}
This quantity defines a metric on the state space of the open quantum system. It is contractive under completely positive and trace-preserving maps, and it is finite for all pairs of reduced density operators.

Let $\Phi(t,0)$ denote the reduced dynamical map and let
\begin{equation}
\rho_S^{1,2}(t)=\Phi(t,0)\rho_S^{1,2}(0)
\end{equation}
be the corresponding evolved states. For Markovian dynamics, contractivity implies that $D\bigl(\rho_S^1(t),\rho_S^2(t)\bigr)$ decreases monotonically for every pair of initial states. Non-Markovianity is instead associated with time intervals in which the trace distance increases. Introducing
\begin{equation}
\sigma\bigl(t,\rho_S^{1,2}(0)\bigr)
=
\frac{\mathrm{d}}{\mathrm{d}t}
D\bigl(\rho_S^{1}(t),\rho_S^{2}(t)\bigr),
\end{equation}
one identifies non-Markovian behavior whenever $\sigma\bigl(t,\rho_S^{1,2}(0)\bigr)>0$. 

The BLP measure quantifies the total recovery of distinguishability during the evolution~\cite{breuer_measure_2009}:
\begin{equation}
\mathcal{N}(\Phi)
=
\max_{\rho_S^{1,2}(0)}
\int_{\sigma>0}\mathrm{d}t\,
\sigma\bigl(t,\rho_S^{1,2}(0)\bigr).
\label{eq:BLPdefinition}
\end{equation}
The integral is taken over all intervals in which $\sigma$ is positive, and the maximization is performed over all pairs of initial system states.

In our calculations, for fixed circuit parameters and for an initial thermal state of the transmission line at temperature $T$, we propagate the reduced dynamics with HEOM over a finite time window of $\omega_qt = 100$. For each candidate pair of initial states $\rho_S^{1,2}(0)$, we evaluate
\begin{equation}
D(t)=D\bigl(\rho_S^1(t),\rho_S^2(t)\bigr),
\end{equation}
identify the intervals where $\dot{D}(t)>0$, and sum the corresponding net increases. The BLP measure is then estimated by maximizing this quantity over a Monte Carlo ensemble of admissible initial pairs~\cite{clos_quantification_2012}.

For a two-qubit system, the search can be parameterized conveniently in terms of $\Delta\rho=\rho^1-\rho^2$.
Since $\operatorname{tr}(\Delta\rho)=0$, the operator $\Delta\rho$ can be expanded in the Pauli-product basis as
\begin{equation}
\Delta \rho
=
\frac{1}{4}\left(
\sum_{i=1}^3 \tilde{r}_i\, \hat{\sigma}_i \otimes \hat{\mathbb{I}}
+
\sum_{j=1}^3 \tilde{s}_j\, \hat{\mathbb{I}} \otimes \hat{\sigma}_j
+
\sum_{i,j=1}^3 \tilde{t}_{ij}\, \hat{\sigma}_i \otimes \hat{\sigma}_j
\right),
\label{eq:delta_rho_pauli}
\end{equation}
where the coefficients $\tilde{r}_i$, $\tilde{s}_j$, and $\tilde{t}_{ij}$ are given by Hilbert--Schmidt projections.

The vectors $\tilde{\mathbf{r}}, \tilde{\mathbf{s}}\in\mathbb{R}^3$ represent the local Bloch components of $\Delta\rho$, while $\tilde{T}=(\tilde{t}_{ij})\in\mathbb{R}^{3\times 3}$ is the two-qubit correlation tensor. Together, Eq.~\eqref{eq:delta_rho_pauli} contains $15$ real parameters, as expected for the space of traceless Hermitian operators in $\mathbb{C}^2\otimes\mathbb{C}^2$.

In the Monte Carlo maximization, we sample $4096$ admissible pairs $\rho^{1,2}(0)$, equivalently admissible $\Delta\rho$, subject to positivity and unit trace of each density matrix, propagate both states under the same dynamical map $\Phi(t,0)$, and retain the largest value of Eq.~\eqref{eq:BLPdefinition} over the sampled ensemble.

\begin{table*}
  \centering
  \scriptsize
  \setlength{\tabcolsep}{3pt}
  \renewcommand{\arraystretch}{1.15}
  \caption{ Initial two-qubit density matrices used for the representative dynamics shown in Fig.~\ref{fig:LongLineContinuum}(c)--(g). For each parameter point, the reported density matrix is one member of the optimal pair obtained from the BLP maximization procedure. The \(4\times4\) density matrices are parameterized as in Eq.~\eqref{eq:bloch_two_qubit_table}; the table entries report the local Bloch vectors $r_i$ and $s_j$, together with the correlation tensor components $t_{ij}$.}
  \label{tab:bloch}
  \resizebox{\textwidth}{!}{%
  \begin{tabular}{c ccc ccc ccccccccc}
    \toprule
    & \multicolumn{3}{c}{$r_i$}
    & \multicolumn{3}{c}{$s_j$}
    & \multicolumn{9}{c}{$t_{ij}$} \\
    \cmidrule(lr){2-4}
    \cmidrule(lr){5-7}
    \cmidrule(lr){8-16}
    & $r_1$ & $r_2$ & $r_3$
    & $s_1$ & $s_2$ & $s_3$
    & $t_{11}$ & $t_{12}$ & $t_{13}$
    & $t_{21}$ & $t_{22}$ & $t_{23}$
    & $t_{31}$ & $t_{32}$ & $t_{33}$ \\
    \midrule
    $(c)$
      & $-0.465$ & $0.217$ & $0.465$
      & $0.465$  & $0.000$ & $0.000$
      & $-0.433$ & $-0.465$ & $0.217$
      & $0.465$  & $0.000$ & $0.465$
      & $0.433$  & $-0.465$ & $-0.217$ \\
    $(d)$
      & $0.233$  & $0.233$  & $0.136$
      & $0.233$  & $0.233$  & $-0.136$
      & $-0.368$ & $0.136$  & $-0.504$
      & $0.136$  & $-0.504$ & $0.000$
      & $0.504$  & $0.000$  & $-0.504$ \\
    $(e)$
      & $-0.812$ & $-0.220$ & $-0.188$
      & $-0.408$ & $-0.624$ & $0.435$
      & $0.404$  & $0.596$  & $-0.596$
      & $0.377$  & $0.216$  & $0.220$
      & $-0.188$ & $0.435$  & $0.027$ \\
    $(f)$
      & $0.610$  & $0.485$  & $0.295$
      & $0.644$  & $0.485$  & $0.114$
      & $0.644$  & $0.212$  & $-0.114$
      & $0.360$  & $0.295$  & $0.360$
      & $0.190$  & $0.485$  & $-0.190$ \\
    $(g)$
      & $-0.313$ & $-0.127$ & $0.402$
      & $0.253$  & $0.067$  & $-0.030$
      & $-0.156$ & $0.283$  & $0.000$
      & $0.313$  & $0.313$  & $-0.156$
      & $0.499$  & $0.529$  & $0.186$ \\
    \bottomrule
  \end{tabular}%
  }
\end{table*}

{Table~\ref{tab:bloch} reports the generalized Bloch representation of the \(4\times4\) two-qubit density matrices used as initial states for the representative dynamics shown in Fig.~\ref{fig:LongLineContinuum}(c)--(g). These states are selected from the optimal pairs obtained in the BLP maximization. For each parameter point, the reported state is selected from the optimal
pair obtained in the BLP maximization.
We represent the corresponding density matrix as}
\begin{multline}
\rho_0 =
\frac{1}{4}
\hat{\mathbb{I}}\otimes\hat{\mathbb{I}}
+\frac{1}{4}\sum_i r_i\,\hat{\sigma}_i\otimes\hat{\mathbb{I}} + \\
+\frac{1}{4}\sum_j s_j\,\hat{\mathbb{I}}\otimes\hat{\sigma}_j
+\frac{1}{4}\sum_{ij} t_{ij}\,\hat{\sigma}_i\otimes\hat{\sigma}_j,
\label{eq:bloch_two_qubit_table}
\end{multline}
where $i,j\in\{1,2,3\}$ and $\sigma_{1,2,3}=(\sigma_x,\sigma_y,\sigma_z)$. The vectors $r_i$ and $s_j$ describe the local Bloch components of the two qubits, while $t_{ij}$ are the components of the two-qubit correlation tensor.


\section{Second-order time-convolutionless master equation}
\label{app:TCL2}
In this Appendix, we derive the second-order time-convolutionless master
equation used for the long-line continuum region \cite{breuer_theory_2002}. The
interaction Hamiltonian \eqref{eq:Hint_bosons2} reads
\begin{equation}
\hat H_{\mathrm{int}}
=
\hbar
\sum_{\alpha=\pm}
\hat L_{\alpha}\otimes \hat B_{\alpha},
\end{equation}
where $\alpha = \pm$ labels the even/odd parity sector and $$\hat{B}_{\pm}=\displaystyle\sum_{n\in \, \substack{2,4,\ldots\\ 1,3,\ldots}}^{\infty} g_{n} \left(\hat{b}_{n}+\hat{b}_{n}^{\dagger}\right)$$
In the continuum limit considered in the main text, the even and odd
sectors are represented by two independent reservoirs with identical
correlation functions,
\begin{equation}
\operatorname{Tr}_{E}
\left[
\hat B_{\alpha}(\tau)
\hat B_{\beta}(0)
\hat\rho_E
\right]
=
\delta_{\alpha\beta} C(\tau),
\end{equation}
with $C(-\tau)=C^*(\tau)$. The bath operators are assumed to have a zero mean.

In the interaction picture, the operators $\hat{L}_{\alpha}$ read

\begin{equation}
\hat L_{\alpha}(t)
=
-i e^{-i\omega_q t}\hat J_{\alpha}
+
i e^{i\omega_q t}\hat J_{\alpha}^{\dagger},
\end{equation}

where 

\begin{equation}
\hat J_{\pm}
=
\hat\sigma_-^{(1)}
\pm \hat\sigma_-^{(2)}.
\label{eq:jump_ops}
\end{equation} 

To second order in the system--bath coupling, and for an initially
factorized state, the TCL2 equation in the interaction picture is
\begin{multline}
\frac{d}{dt}\tilde\rho_S(t)
=
-\sum_{\alpha=\pm}
\int_0^t d\tau
\Big\{
C(\tau)
\left[
\hat L_{\alpha}(t),
\hat L_{\alpha}(t-\tau)\tilde\rho_S(t)
\right]
\\
+
C^*(\tau)
\left[
\tilde\rho_S(t)\hat L_{\alpha}(t-\tau),
\hat L_{\alpha}(t)
\right]
\Big\}.
\end{multline}
By defining the emission and absorption coefficients
\begin{equation}
\Gamma_{\downarrow}(t)
=
\int_0^t d\tau\,
e^{i\omega_q \tau} C(\tau),
\quad
\Gamma_{\uparrow}(t)
=
\int_0^t d\tau\,
e^{-i\omega_q \tau} C(\tau),
\end{equation}
the TCL2 equation in Schrödinger picture is
\begin{multline}
\frac{d}{dt}\rho_S(t)
=
-\frac{i}{\hbar}
\left[
\hat H_S+\hat H_{\mathrm{LS}}(t),
\rho_S(t)
\right]
+ \\
\mathcal{D}_{\mathrm{sec}}(t)\rho_S(t)
+
\mathcal{D}_{\mathrm{nsec}}(t)\rho_S(t).
\label{eq:TCL2}
\end{multline}
The Lamb-shift Hamiltonian is given by
\begin{equation}
\hat H_{\mathrm{LS}}(t)
=
\hbar
\left[
\operatorname{Im}\Gamma_{\downarrow}(t)
-
\operatorname{Im}\Gamma_{\uparrow}(t)
\right]
\left(
\hat\sigma_z^{(1)}
+
\hat\sigma_z^{(2)}
\right).
\end{equation}
The secular
contribution is given by
\begin{equation}
\mathcal{D}_{\mathrm{sec}}(t)\rho_S
=
\gamma_{\downarrow}(t)
\sum_{j=1}^{2}
\mathcal{D}[\hat\sigma_-^{(j)}]\rho_S
+
\gamma_{\uparrow}(t)
\sum_{j=1}^{2}
\mathcal{D}[\hat\sigma_+^{(j)}]\rho_S,
\end{equation}
where $\gamma_{a}(t)=4\,\operatorname{Re}\Gamma_{a}(t)$ are the local rates with 
$a = \downarrow\, , \uparrow$, while the canonical dissipator is $\mathcal{D}[\hat O]\rho=\hat O\rho \hat O^\dagger-
\frac{1}{2}\left\{\hat O^\dagger \hat O,\rho \right\}.
$
The prefactors in the local rates follow from the sum over the two
independent parity sectors, each with the same correlation function.
The nonsecular contribution is given by
\begin{equation}
\mathcal{D}_{\mathrm{nsec}}(t)\rho_S
=
\chi(t)
\sum_{j=1}^{2}
\hat\sigma_-^{(j)}\rho_S\hat\sigma_-^{(j)}
+
\chi^*(t)
\sum_{j=1}^{2}
\hat\sigma_+^{(j)}\rho_S\hat\sigma_+^{(j)},
\end{equation}
with $\chi(t)=-2 e^{-2i\omega_q t}\left[\Gamma_{\downarrow}(t)+\Gamma_{\uparrow}^*(t)\right]$.

The resulting TCL2 generator is local in the two qubits. That is a consequence of the equal statistical weight of the two independent
parity sectors in the symmetric continuum limit. Indeed, when the
contributions of $\hat J_{\pm}$ are summed, all the cross-qubit terms vanish. Therefore, the continuum TCL2 equation
contains local relaxation, excitation, Lamb-shift, and nonsecular terms,
but no bath-induced dissipative coupling between the two qubits.

\section{GKLS equation}
\label{app:GKLSeq}

In this Appendix, we derive the Gorini--Kossakowski--Lindblad–Sudarshan (GKLS) equation used as a Markovian benchmark in the long-line continuum
description. We use the same notation and conventions introduced in
Appendix~\ref{app:TCL2}.
The GKLS equation is obtained from the TCL2 generator by extending the
upper integration limit to infinity and retaining only the secular terms.
We introduce the Markovian one-sided coefficients
\begin{equation}
\Xi_{\downarrow}
=
\int_0^\infty d\tau\,
e^{i\omega_q \tau}C(\tau),
\qquad
\Xi_{\uparrow}
=
\int_0^\infty d\tau\,
e^{-i\omega_q \tau}C(\tau).
\end{equation}
Their real parts define the emission and absorption rates of a single
parity channel,
\begin{equation}
\kappa_{\downarrow}
=
2\,\operatorname{Re}\Xi_{\downarrow},
\qquad
\kappa_{\uparrow}
=
2\,\operatorname{Re}\Xi_{\uparrow}.
\end{equation}
With the spectral-density convention of Eq.~\eqref{eq:spec_density_conv}, these
rates are
\begin{equation}
\kappa_{\downarrow}
=
2J(\omega_q)\left[\bar n(\omega_q)+1\right],
\qquad
\kappa_{\uparrow}
=
2J(\omega_q)\bar n(\omega_q),
\label{eq:rates_GKLS}
\end{equation}
where $\bar n(\omega)
=
\left(e^{\beta\hbar\omega}-1\right)^{-1}$.

The resulting GKLS master equation can first be written in the parity-channel
basis as
\begin{multline}
\dot\rho
=
-\frac{i}{\hbar}
\left[
\hat H_S+\hat H_{\mathrm{LS}},
\rho
\right]
\\
+
\kappa_{\downarrow}
\sum_{\alpha=\pm}
\mathcal D[\hat J_{\alpha}]\rho
+
\kappa_{\uparrow}
\sum_{\alpha=\pm}
\mathcal D[\hat J_{\alpha}^{\dagger}]\rho.
\label{eq:GKLS_master_parity}
\end{multline}

In our case, due to the symmetry, the two parity channels have identical
rates; therefore, Eq.~\eqref{eq:GKLS_master_parity} takes the local form
\begin{multline}
\dot\rho
=
-\frac{i}{\hbar}
\left[
\hat H_S+\hat H_{\mathrm{LS}},
\rho
\right]
\\
+
\gamma_{\downarrow}
\sum_{j=1}^{2}
\mathcal D[\hat\sigma_-^{(j)}]\rho
+
\gamma_{\uparrow}
\sum_{j=1}^{2}
\mathcal D[\hat\sigma_+^{(j)}]\rho ,
\label{eq:GKLS_master_local}
\end{multline}
with local rates $\gamma_{\downarrow}
=
2\kappa_{\downarrow}$, $\gamma_{\uparrow}
=
2\kappa_{\uparrow}$.
Thus, the GKLS dissipator is local in the two qubits and it is the Markovian
counterpart of the locality of the secular TCL2 contribution derived in
Appendix~\ref{app:TCL2}.

The Lamb-shift Hamiltonian is determined by the imaginary parts of the same
one-sided Markovian coefficients. Equivalently, it can be written as
\begin{equation}
\hat H_{\mathrm{LS}}
=
\hbar
\sum_{\alpha=\pm}
\left[
S(\omega_q)
\hat J_{\alpha}^{\dagger}\hat J_{\alpha}
+
S(-\omega_q)
\hat J_{\alpha}\hat J_{\alpha}^{\dagger}
\right],
\label{eq:HLS_GKLS_parity}
\end{equation}
where
\begin{equation}
S(\omega)
=
\mathcal P
\int_{0}^{\infty}
d\omega'\,
\frac{J(\omega')}{\pi}
\left[
\frac{\bar n(\omega')+1}{\omega-\omega'}
+
\frac{\bar n(\omega')}{\omega+\omega'}
\right].
\label{eq:Somega_def}
\end{equation}
By computing the jump operators, one obtains, up to an irrelevant constant energy shift,
\begin{equation}
\hat H_{\mathrm{LS}}
\simeq
\hbar
\left[
S(\omega_q)-S(-\omega_q)
\right]
\left(
\hat\sigma_z^{(1)}
+
\hat\sigma_z^{(2)}
\right).
\label{eq:HLS_GKLS_local}
\end{equation}


\clearpage
\onecolumngrid               

\begin{center}
  \textbf{\large Supplemental Material}
\end{center}

\setcounter{section}{0}
\setcounter{equation}{0}
\setcounter{figure}{0}
\setcounter{table}{0}
\renewcommand{\thesection}{S\arabic{section}}
\renewcommand{\theequation}{S\arabic{equation}}
\renewcommand{\thefigure}{S\arabic{figure}}
\renewcommand{\thetable}{S\arabic{table}}

\section{Connection with the full impedance matrix}
\label{sec:Impedance}

Here, we relate the normal-mode construction used in the main text to the black-box impedance of the corresponding linear two-port network seen by the two Josephson junctions \cite{solgun_blackbox_2014,solgun_simple_2019,labarca_toolbox_2024}. For the symmetric circuit in 
Fig.~\ref{fig:finiteTLoverview}, the admittance matrix $\boldsymbol{Y}$ in the Laplace variable $s$,  seen from the two qubit ports is
\begin{equation}
\boldsymbol{Y}(s)
=
\left[
\boldsymbol{Z}_{\mathrm{TL}}(s)
+
\frac{1}{s C_c}\boldsymbol{1}_{2}
\right]^{-1}
+
s C_J \,\boldsymbol{1}_{2}.
\end{equation}
The corresponding port impedance is
\begin{equation}
\boldsymbol{Z}(s)=\boldsymbol{Y}^{-1}(s).
\end{equation}
The two-port impedance matrix associated with the transmission line alone is
\begin{equation}
\boldsymbol{Z}_{\mathrm{TL}}(s)
=
Z_{\mathrm{TL}}
\begin{pmatrix}
\operatorname{coth}\left(s d/v_p\right) &
\operatorname{csch}\left(s d/v_p\right)
\\
\operatorname{csch}\left(s d/v_p\right) &
\operatorname{coth}\left(s d/v_p\right)
\end{pmatrix}.
\end{equation}

The parity transformation, corresponding to symmetric and antisymmetric excitations
\begin{equation}
\boldsymbol{P}=
\frac{1}{\sqrt{2}}
\begin{pmatrix}
1 & 1
\\
1 & -1
\end{pmatrix}
\end{equation}
diagonalizes the line impedance. In this basis,
\begin{equation}
\boldsymbol{P}^\intercal \boldsymbol{Z}_{\mathrm{TL}}(s) \boldsymbol{P}
=
\text{diag} \left(
z_{+}(s),  z_{-}(s) \right)
\end{equation}
with
\begin{equation}
z_{\pm}(s)
=
Z_{\mathrm{TL}}
\begin{array}{c}
\operatorname{coth}\\
\operatorname{tanh}
\end{array}
\left(\frac{s d}{2v_p}\right),
\end{equation}
Here \(+\) and \(-\) are again associated with the even and odd sectors, respectively.

The Foster expansion of the impedance $\boldsymbol{Z}(s)$ has the expression
\begin{equation}
\boldsymbol{Z}(s)=\frac{\boldsymbol{A}_{0}}{s}+\sum_{n} \frac{s \boldsymbol{A}_{n}}{s^{2}+\omega_{n}^{2}}.
\label{eq:Zfoster}
\end{equation}
The zero-frequency term \( \boldsymbol{A}_0/s \) represents the static capacitive part of the chosen black-box partition. The dynamical transmission line modes correspond to the finite-frequency residues \( \boldsymbol{A}_n \). Therefore, if the full impedance \( \boldsymbol{Z}(s) \) is used as the black-box object, the static capacitance associated with \( \boldsymbol{A}_0 \) should not be counted again elsewhere in the Hamiltonian.

The expressions of the scalar impedances on the diagonal of $\mathbf{Z}$ after the parity transformation are
\begin{equation}
Z_{\sigma}(s)
=
\frac{1+s C_c z_{\sigma}(s)}
{s C_{\Sigma}\left[1+s C_s z_{\sigma}(s)\right]}
\end{equation}
where $\sigma=\pm$ and $C_{\Sigma}=C_J+C_c$. 
The poles of $\boldsymbol{Z}(s)$ are therefore fixed by
\begin{equation}
1+s C_s z_{\sigma}(s)=0 .
\end{equation}
Setting \(s=i\omega_n\) with \(k_n=\omega_n/v_p\), this equation gives the equations \eqref{eq:odd_eig} and \eqref{eq:even_eig} obtained from the auxiliary
eigenvalue problem.

The corresponding mode coupling appearing in the Hamiltonian is obtained from the diagonal residue as
\begin{equation}
g_n^2
=
\frac{Q_T^2\omega_n}{2\hbar}
\left(\boldsymbol{A}_{n}\right)_{11}.
\end{equation}
Equivalently, the off-diagonal residue gives
\begin{equation}
\frac{Q_T^2\omega_n}{2\hbar}
\left(\boldsymbol{A}_{n}\right)_{12}
=
(-1)^n g_n^2 .
\end{equation}
Therefore, the diagonal residues determine the local Lamb-shift weights, while
the off-diagonal residues determine the parity-weighted qubit--qubit exchange.

These results show that the normal-mode Hamiltonian and the full impedance-matrix
description are two equivalent representations of the same linear circuit.
\end{document}